\documentclass[10pt]{article}
\usepackage{amsmath,amsthm,latexsym,amssymb,amsfonts,epsfig}

\makeatletter
\@addtoreset{equation}{section}
\makeatother

\def\be{\begin{equation}}
\def\ee{\end{equation}}
\def\ba{\begin{eqnarray}}
\def\ea{\end{eqnarray}}

\newcommand\nn{\nonumber}
\newcommand\q{\quad}
\def\Nl{{\mathchoice
{\setbox0=\hbox{$\displaystyle\rm N$}\hbox{\hbox to0pt
{\kern0.4\wd0\vrule height0.9\ht0\hss}\box0}}
{\setbox0=\hbox{$\textstyle\rm N$}\hbox{\hbox to0pt
{\kern0.4\wd0\vrule height0.9\ht0\hss}\box0}}
{\setbox0=\hbox{$\scriptstyle\rm N$}\hbox{\hbox to0pt
{\kern0.4\wd0\vrule height0.9\ht0\hss}\box0}}
{\setbox0=\hbox{$\scriptscriptstyle\rm N$}\hbox{\hbox to0pt
{\kern0.4\wd0\vrule height0.9\ht0\hss}\box0}}}}
\def\Zl{{\mathchoice
{\setbox0=\hbox{$\displaystyle\rm Z$}\hbox{\hbox to0pt
{\kern0.4\wd0\vrule height0.9\ht0\hss}\box0}}
{\setbox0=\hbox{$\textstyle\rm Z$}\hbox{\hbox to0pt
{\kern0.4\wd0\vrule height0.9\ht0\hss}\box0}}
{\setbox0=\hbox{$\scriptstyle\rm Z$}\hbox{\hbox to0pt
{\kern0.4\wd0\vrule height0.9\ht0\hss}\box0}}
{\setbox0=\hbox{$\scriptscriptstyle\rm Z$}\hbox{\hbox to0pt
{\kern0.4\wd0\vrule height0.9\ht0\hss}\box0}}}}
\def\Ql{{\mathchoice
{\setbox0=\hbox{$\displaystyle\rm Q$}\hbox{\hbox to0pt
{\kern0.4\wd0\vrule height0.9\ht0\hss}\box0}}
{\setbox0=\hbox{$\textstyle\rm Q$}\hbox{\hbox to0pt
{\kern0.4\wd0\vrule height0.9\ht0\hss}\box0}}
{\setbox0=\hbox{$\scriptstyle\rm Q$}\hbox{\hbox to0pt
{\kern0.4\wd0\vrule height0.9\ht0\hss}\box0}}
{\setbox0=\hbox{$\scriptscriptstyle\rm Q$}\hbox{\hbox to0pt
{\kern0.4\wd0\vrule height0.9\ht0\hss}\box0}}}}
\def\Rl{{\mathchoice
{\setbox0=\hbox{$\displaystyle\rm R$}\hbox{\hbox to0pt
{\kern0.4\wd0\vrule height0.9\ht0\hss}\box0}}
{\setbox0=\hbox{$\textstyle\rm R$}\hbox{\hbox to0pt
{\kern0.4\wd0\vrule height0.9\ht0\hss}\box0}}
{\setbox0=\hbox{$\scriptstyle\rm R$}\hbox{\hbox to0pt
{\kern0.4\wd0\vrule height0.9\ht0\hss}\box0}}
{\setbox0=\hbox{$\scriptscriptstyle\rm R$}\hbox{\hbox to0pt
{\kern0.4\wd0\vrule height0.9\ht0\hss}\box0}}}}
\def\Cl{{\mathchoice
{\setbox0=\hbox{$\displaystyle\rm C$}\hbox{\hbox to0pt
{\kern0.4\wd0\vrule height0.9\ht0\hss}\box0}}
{\setbox0=\hbox{$\textstyle\rm C$}\hbox{\hbox to0pt
{\kern0.4\wd0\vrule height0.9\ht0\hss}\box0}}
{\setbox0=\hbox{$\scriptstyle\rm C$}\hbox{\hbox to0pt
{\kern0.4\wd0\vrule height0.9\ht0\hss}\box0}}
{\setbox0=\hbox{$\scriptscriptstyle\rm C$}\hbox{\hbox to0pt
{\kern0.4\wd0\vrule height0.9\ht0\hss}\box0}}}}
\def\Hl{{\mathchoice
{\setbox0=\hbox{$\displaystyle\rm H$}\hbox{\hbox to0pt
{\kern0.4\wd0\vrule height0.9\ht0\hss}\box0}}
{\setbox0=\hbox{$\textstyle\rm H$}\hbox{\hbox to0pt
{\kern0.4\wd0\vrule height0.9\ht0\hss}\box0}}
{\setbox0=\hbox{$\scriptstyle\rm H$}\hbox{\hbox to0pt
{\kern0.4\wd0\vrule height0.9\ht0\hss}\box0}}
{\setbox0=\hbox{$\scriptscriptstyle\rm H$}\hbox{\hbox to0pt
{\kern0.4\wd0\vrule height0.9\ht0\hss}\box0}}}}
\def\Ol{{\mathchoice
{\setbox0=\hbox{$\displaystyle\rm O$}\hbox{\hbox to0pt
{\kern0.4\wd0\vrule height0.9\ht0\hss}\box0}}
{\setbox0=\hbox{$\textstyle\rm O$}\hbox{\hbox to0pt
{\kern0.4\wd0\vrule height0.9\ht0\hss}\box0}}
{\setbox0=\hbox{$\scriptstyle\rm O$}\hbox{\hbox to0pt
{\kern0.4\wd0\vrule height0.9\ht0\hss}\box0}}
{\setbox0=\hbox{$\scriptscriptstyle\rm O$}\hbox{\hbox to0pt
{\kern0.4\wd0\vrule height0.9\ht0\hss}\box0}}}}
\newcommand{\bd}{\mathbf d}

\begin{document}

\title{A perturbative approach to Dirac observables and their space--time algebra}
\author{Bianca Dittrich\thanks{bdittrich@perimeterinstitute.ca} \\
\it \small Perimeter Institute for Theoretical Physics, 
31 Caroline Street North, Waterloo, ON N2L 2Y5, Canada \\
 Johannes Tambornino\thanks{jtambornino@perimeterinstitute.ca}\\
\it \small Perimeter Institute for Theoretical Physics,
31 Caroline Street North, Waterloo, ON N2L 2Y5, Canada \\
\it \small and\\
\it \small Institut f\"ur Physik, RWTH Aachen, D-52056 Aachen, Germany
}


\maketitle

\begin{abstract}
We introduce a general approximation scheme in order to calculate gauge invariant observables in the canonical formulation of general relativity. Using this scheme we will show how the observables and the dynamics of field theories on a fixed background or equivalently the observables of the linearized theory can be understood as an approximation to the observables in full general relativity. Gauge invariant corrections can be calculated up to an arbitrary high order and we will explicitly calculate the first non--trivial correction. 

Furthermore we will make a first investigation into the Poisson algebra between observables corresponding to fields at different space--time points and consider the locality properties of the observables.

\end{abstract}

\section{Introduction}\label{intro}

One of the most important issues for a quantum theory of gravity is the construction and interpretation of the observables of the theory. We will address this issue in the framework of the canonical theory, where the gauge independent observables are called Dirac observables. So far  -- apart from the ADM charges \cite{ADM} in the context of asymptotically flat space--times -- no Dirac observables are known explicitly for pure gravity. Moreover in \cite{torre} it is shown that with the exception of the ADM charges, Dirac observables have to include an infinite number of spatial derivatives. Hence we expect that it will be tremendously difficult to calculate Dirac observables and that the only resort may be to develop approximation methods for Dirac observables in order to make physical predictions. One proposal \cite{gambini} for an approximation scheme is an expansion in the inverse cosmological constant. 

In this work we will propose a general approximation scheme for Dirac observables, based on the concept of complete observables \cite{rovelli}. A complete observable $F_{[f,T]}(\tau)$ is a special kind of relational observable: Using some of the dynamical variables $T$ as clocks, the complete observable gives the value of some other dynamical variable $f$ at that instant at which the clock variables assume certain values $\tau$. 

The complete observables can be computed using a power series \cite{bd1,bd2} in the (phase space dependent) clock variables. This power series is a natural starting point for a perturbative approach and indeed we will use it to obtain an approximation scheme by expanding the series in the ``fluctuations'' around some fixed phase space point.

We will apply this approximation scheme to general relativity. In doing so, we have to make certain choices -- the most important being the choice of the clock variables. Here our guide line is that we want to have a good approximation to the observables of field theory on a fixed background which in this work will be the flat Minkowski background. As we will see this results in observables which in the zero gravity limit (i.e. for $\kappa=0$ and vanishing gravitational fluctuations) coincide with the usual observables of field theory on a flat background. The gravity corrections can be calculated explicitly order by order and are connected to the standard perturbation theory. The first order observables are given by the observables of linearized general relativity, hence this method gives us a precise understanding of the observables of the linearized theory, for instance the graviton, as approximations to observables of the full theory. Moreover the approximation scheme in this work gives a precise proposal how to compute higher order corrections to the observables of the linearized theory. These higher order corrections are in a consistent way gauge invariant to a certain order -- to make these corrections completely gauge invariant one would have to add terms which are of higher order than the corrections themselves. One important point is that we do not only manage to approximate well the fields at one specific time, but obtain also an approximation to the fields at arbitrary times, which makes the extraction of dynamical information much easier. 

This brings us to the second issue mentioned above, namely the interpretation of the observables, in particular in view of the quantum theory. One question for instance is, whether it is possible at all to construct Dirac observables which give the standard (local) field observables in some limit and what kind of locality properties these observables have. Having a (perturbative) computation scheme for Dirac observables at hand allows one to examine these issues in more detail. 

Deviations from the standard field observables could result in fundamental uncertainties for the observables in a quantum theory including gravity, as is argued in \cite{hartle} using relational observables. One crucial aspect for these deviations is that if one uses relational observables one has to specify some of the dynamical fields as clock variables. However these clock variables are dynamical. Since measurement involves always a disturbance of the system this means that one has also to expect a disturbance of the clock variables if one would measure a complete observable. Another aspect is, that if we use matter fields as clocks these matter fields are coupled to the gravitational field and therefore influence all the other fields. In the classical theory these influences will show up for example in the Poisson algebra of Dirac observables corresponding to measurements at different space--time points. Here it is important to consider space--time points at different times, since equal--time Poisson brackets will be zero for the cases we are interested in. The Poisson algebra will be reflected in uncertainty relations in the quantum theory for these measurements. Connected to this is the question whether this Poisson algebra satisfies any locality properties, for instance whether fields at spatially separated points Poisson commute. In this work we will make a first investigation into the classical Poisson algebra of observables corresponding to measurements at different space--time points.  

However the Poisson algebra and the uncertainty relations which follow from this algebra will depend on the choice of clock variables and it is important to understand in which ways this choice matters. Let us consider a very simple example, namely two parametrized particles with the Hamiltonian constraint
\ba\label{in1}            
C=p_t+\frac{p_1^2}{2 m_1}+\frac{p_2^2}{2 m_2} 
\ea
where $p_t$ is the conjugated momentum to the time variable $t$ and $p_1,p_2$ are conjugated to the two position variables $q_1,q_2$. A natural choice for a clock variable is $t$ and we can ask for the position of the first particle at that moment at which $t$ assumes the value $\tau$. We will denote this observable by $F_{[q_1;t]}(\tau)$ and it can be easily calculated to be
\ba\label{in2}
F_{[q_1;t]}(\tau)=q_1+\frac{p_1}{m}(\tau-t)  \q .
\ea
It Poisson commutes with the constraint and is therefore a Dirac observable. The Poisson bracket of two observables $F_{[q_1;t]}(\tau_1)$ and $F_{[q_1;t]}(\tau_2)$ at two different clock values $\tau_1$ and $\tau_2$ is phase space independent
\ba\label{in3}
\{F_{[q_1;t]}(\tau_1),F_{[q_1;t]}(\tau_2)\}=\frac{1}{m_1}(\tau_2-\tau_1) \q .
\ea

Now one could also choose the position of the second particle as a clock variable and ask for the position of the first particle at that moment at which the second particle has position $\tau'$. The corresponding observable is
\ba\label{in4}
F_{[q_1;q_2]}(\tau')=q_1+\frac{p_1}{m_1}\frac{m_2}{p_2}(\tau'-q_2)  \q .
\ea
If one ignores that $\tau$ and $\tau'$ refer to different clocks, (\ref{in4}) looks of course quite different from (\ref{in3}). However if one takes into account that the value $\tau$ is reached at that moment at which 
\ba\label{in4a}
\tau'=F_{[q_2;t]}=q_2+\frac{p_2}{m_2}(\tau-t)
\ea
and uses this to replace $\tau'$ in (\ref{in4}) one will get back to (\ref{in3}). In so far both choices of clock variables give us the same time evolution if one takes into account the ``translation'' (\ref{in4a}) between the clock readings $\tau$ and $\tau'$.  This is only a classical consideration, the quantization of (\ref{in4}) will have quite different properties from the standard position operator due to the difficulties involved in quantizing the inverse momentum $p_2$ in (\ref{in4}), see \cite{rov2,unruh} and references therein.

However the Poisson bracket between two observables with respect to the clock variable $q_2$ is given by
\ba\label{in5}
\{F_{[q_1;q_2]}(\tau'_1),F_{[q_1;q_2]}(\tau'_2)\}=\big[ \frac{1}{m_1}(\tau'_2-\tau'_1)\big]\big[\frac{m_2}{p_2}\big]\big[ 1+\frac{p_1^2}{m_1}\frac{m_2}{p_2^2}\big] \q .
\ea
The first factor in square brackets on the right hand side of (\ref{in5}) is equal to our previous result (\ref{in3}), the second factor is due to the translation between $\tau$-- and $\tau'$--parameters. The third factor can be seen as a correction term resulting from the use of an unusual clock variable. The correction is proportional to the kinetic energy of the particle observed divided by the kinetic energy of the particle used as a clock. In the limit of large energy of the particle used as clock compared to the observed particle we get back to the previous result (\ref{in3}). This corresponds to the intuition that the second particle is a good clock if its velocity is large, i.e. if it has large energy. Interestingly, if one attempts to quantize operators of the kind (\ref{in4}) one derives an inherent uncertainty inverse to the kinetic energy of the particle used as clock \cite{unruh}. To take the limit of large energy for the clock variables is problematic in general relativity because of backreaction and ultimately black hole formation.

We see that different choices for the clock variables lead to different predictions for the uncertainties for the complete observables. For general relativity the question arises what kind of clock variables are available. For matter clocks and clocks built from curvature scalars we expect a behaviour similar to (\ref{in5}). These clocks have to have a certain energy in order to be good clocks and therefore it is in question whether the limit to flat space can be performed. The question arises whether there are any ``good'' clocks near flat space built from the gravitational degrees of freedom and what properties the Poisson algebra of the corresponding complete observables has.

\subsection{Overview}

In section \ref{abel} we will explain how given a first class constraint set, we can construct new constraints with a different constraint algebra: The Poisson brackets between these new constraints are at least quadratic in the constraints. These new constraints are needed for an explicit power series for the complete observables. We will also introduce an iterative method to obtain a constraint set which is Abelian. These Abelian constraints will simplify calculations for the case of general relativity and facilitate the interpretation of the resulting formula.  

The power series for complete observables serves as a basis for our approximation scheme, which we will define in section \ref{appdirac}. The basic idea is to keep in the power series only terms up to a certain order $q$ in the fluctuations around a certain ``background'' phase space point. For special values of the clock parameters $\tau$ this will result in only finitely many terms in the power series.

In the next section \ref{gr} we summarize some material necessary to apply this approximation scheme to general relativity and in particular we define our clock variables. In \ref{dyn} we will find that this approximation scheme also allows one to compute complete observables corresponding to a one--parameter family of clock parameters, that is to different times. This facilitates the dynamical interpretation of the complete observables. 

Next we will consider the second order approximation in \ref{approximation} and give explicit formulae for the terms one has to calculate. These calculations are performed for a Klein--Gordon field coupled to gravity in \ref{scalar}. The Poisson algebra between the resulting approximate observables is considered in section \ref{poisalg}.

\section{Perturbative Abelianization of the constraints}\label{abel}

In this section we will collect some facts about first class constrained systems that will be important later on. More concretely we will show how one can redefine the constraints such that these new constraints have particular nice properties, which are key for the construction of the complete observables. In the second part we will introduce a method to obtain an Abelian set of constraints.

To simplify notation we will restrict ourselves to a finite dimensional phase space, however the results are easily generalizable to an infinite dimensional phase space.

Let $\{C_j\}_{j=1}^{m}$ be a set of independent\footnote{i.e. the equations $C_j=0$ should be independent from each other} first class constraints on a phase space described by canonical coordinates $\{X^a\}_{a=1}^{2n}$. Choose a phase space point  $X_0:=\{X^a_0\}_{a=1}^{2n}$ satisfying the constraints and introduce new canonical variables as ``fluctuations'' $x^a:=X^a-X^a_0$ around the phase space point. Rewrite the constraints in this new variables and define the first order constraints ${}^{(1)}C_j$ as the part of the constraints linear in the fluctuation variables. We have to assume that these linearized constraints are also independent from each other, in particular they should not vanish. 

Now the linearized constraints are Abelian \cite{lee}
\ba\label{B1}
\{{}^{(1)}C_j,{}^{(1)}C_k\}=0
\ea
as can be seen by expanding the full constraint algebra
\ba\label{B2}
\{{}^{(1)}C_j+{}^{(2)}C_j+\ldots ,\,\, {}^{(1)}C_k+{}^{(2)}C_k+\ldots\}=({}^{(0)}f_{jk}^l+{}^{(1)}f_{jk}^l+\ldots)({}^{(1)}C_l+{}^{(2)}C_l+\ldots )
\ea
order by order. Here the superscript ${}^{(m)}$ left to a symbol denotes its $m$--th order. Hence one can find a set of clock variables $\{T^K\}_{K=1}^m$ canonically conjugated to the linearized constraints, i.e. such that
\ba\label{B3} 
\{T^K,{}^{(1)}C_j\}=\delta^K_j     \q .
\ea
This set of variables can be extended to a canonical coordinate system\footnote{In general this canonical coordinate system is only a local one, i.e. it can be only defined in some region around $X_0$.} by finding an additional set of first order quantities $\{Q^d\}_{d=1}^{n-m}$ and $\{P_d\}_{d=1}^{n-m}$ having vanishing Poisson bracket with the linearized constraints and with the clock variables and satisfying $\{Q^d,P_e\}=\delta^d_e$. 

Define the phase space dependent matrix
\ba\label{B4}
A^K_j:=\{T^K,C_j\}=\delta^K_j+B^K_j
\ea
where $B^K_j$ is at least a first order quantity. Because the zeroth order of $A^K_j$ is given by the identity matrix we can invert $A^K_j$ order by order. We will assume that $A^K_j$ is invertible in some phase phase space region around $X_0$, i.e. that the series defining the perturbative inverse of $A^K_j$ converges in this region. In the following we will restrict our considerations to this region of the phase space.

Using the inverse of $A^K_j$ we define a new set of constraints (equivalent to the old one at all phase space points where the inverse can be defined) by 
\ba\label{B5}
\tilde C_K = (A^{-1})^j_K C_j   \q .
\ea
This new set has the following properties: First it is easy to see that
\ba\label{B6}
\{T^K,\tilde C_M\}=\delta^K_M+\lambda^{KN}_{M}\tilde C_N \simeq \delta^K_M
\ea
where $\lambda^{KN}_{M}=\{T^K,(A^{-1})^j_M\}A^N_j $ are a set of phase space functions and the sign $\simeq$ means weakly equal, that is equal up to terms vanishing on the constraint hypersurface.

From equation (\ref{B6}) one can prove that the new constraints are ``weakly abelian'', i.e. they Poisson commute up to terms quadratic in the constraints: Compute $\{\{T^K,\tilde C_M\},\tilde C_N\}$ first directly and then using the Jacobi identity. Comparing the two results one can conclude that the structure functions $\tilde f_{KJ}^M$ defined by $\{\tilde C_K, \tilde C_J\}=\tilde{f}_{KJ}^M \tilde C_M$ have to vanish on the constraint hypersurface. 

Property (\ref{B6}) of the constraints $\tilde C_K$ will be the key property in order to be able to write down a power series for the complete observables. In the following we will construct  a set of constraints $\check C_K$, which satisfies $\{T^K,\check C_M\}=\delta^K_M$ exactly, i.e. also away from the constraint hypersurface. These constraints will simplify very much the calculations in the later sections.

Note that the flow generated on the constraint hypersurface by a constraint $\tilde C_L$ does not change if we add to $\tilde C_L$ a combination at least quadratic in the constraints
\ba\label{B7} 
\{f,\tilde C_L\}\simeq \{f,\tilde C_L + \mu^{KM}\tilde C_K \tilde C_M\}
\ea
where $\mu^{KM}$ is an arbitrary smooth phase space function.

In particular if we add to the constraints $\tilde C_L$ a combination quadratic or of higher order in the constraints, equation (\ref{B6}) will still hold, the only difference is the exact expression for the phase space functions $\lambda^{KN}_{M}$. However the exact expressions for these coefficients are not important for the proof that the constraints $\tilde C_L$ are weakly abelian.

This can be used in an iterative process to find a set of Abelian constraints (if this process converges). This process can be applied to an arbitrary set of clock variables, as long as the clock variables are Abelian. In the following we will explain the iteration step.

\vspace{0.3cm}

 Assume that one has a set of independent first class constraints $\{\hat C_K\}_{K=1}^m$ and a set of independent phase space functions $\{T^K\}_{K=1}^m$ with the property
\ba\label{B8} 
\{T^K,\hat C_M\}=\delta^K_M +\lambda_M^{K L_1 \cdots L_r} \hat C_{L_1}\cdots \hat C_{L_r}        \q .
\ea
for some $r \geq 1$ and such that
\ba\label{B8a}
\{\hat C_M, \hat C_N\}=\hat f_{MN}^{L_1 \cdots L_{r+1}} \hat C_{L_1}\cdots \hat C_{L_{r+1}}  =O(C^{(r+1)} ) 
\ea
for some phase space functions $\hat f_{MN}^{L_1 \cdots L_{r+1}}$. Here $O(C^1)$ means a smooth function $g$ vanishing on the constraint hypersurface. Such a function $g$ can be written as a combination of the constraints $g=\nu^K \tilde C_K$ with some smooth phase space functions $\nu_k$, see \cite{henneaux}. With $O(C^r)$ we denote a phase space function $g$ which can be written as $g=\nu^{K_1 \cdots K_r}\tilde C_{K_1}\cdots \tilde C_{K_r}$ with some smooth phase space functions $\nu^{K_1 \cdots K_r}$.
 
The $\lambda_M^{K L_1 \cdots L_r}$ in (\ref{B8}) can be a set of arbitrary (smooth in the phase space region of interest) phase space functions, without loss of generality we can assume that they are invariant under permutation of the $\{L_1,\cdots,L_r\}$--indices. Furthermore we will assume that the $\{T^K\}_{K=1}^m$ are Abelian, i.e. have vanishing Poisson brackets among themselves. 

With these assumptions we will show that one can define a new set of constraints 
\ba\label{B9}
{\check C_K}:=\hat C_K + \mu_K^{L_1 \cdots L_{r+1}} \hat C_{L_1}\cdots \hat C_{L_{r+1}}
\ea
with $\mu_K^{L_1 \cdots L_{r+1}}$ symmetric in the $\{L_1,\cdots,L_r\}$--indices such that 
\ba\label{B10}
\{T^K,{\check C_M}\}=\delta^K_M +\lambda_M^{K L_1 \cdots L_r L_{r+1}} \check C_{L_1}\cdots \check C_{L_{r+1}}  \q .   
\ea
Furthermore these new constraints satisfy
\ba\label{B11}
\{ {\check C_K},{\check C_M} \}= {\check f}_{KM}^{L_1 \cdots L_{r+2}}{\check C_{L_1}} \cdots {\check C_{L_{r+2}}}= O(C^{(r+2)})
\ea
for some phase space functions ${\check f}_{KM}^{L_1 \cdots L_{r+2}}$.

\vspace{0.3cm}

{\bf Proof:} First note that we can find the coefficients $\lambda_M^{K L_1 \cdots L_r}$ in (\ref{B8}) at least up to terms proportional to the constraints by taking iterated Poisson brackets between the clock variables and both sides of equation (\ref{B8})  :
\ba\label{B12}
\lambda_M^{K L_1 \cdots L_r}=\frac{1}{r!}\{T^{L_1},\{T^{L_2},\{\cdots,\{T^{L_r},\{T^K,\hat C_M\}\cdots\}  +O(C) \q .
\ea
Hence the $\lambda_M^{K L_1 \cdots L_r}$ are invariant up to $O(C)$--terms under permutation of the $\{K,L_1,\cdots,L_r\}$--indices. (We have assumed that the clock variables are Abelian.) 

Using equation (\ref{B9}) as an ansatz in equation (\ref{B10}) and exploiting the assumption (\ref{B8}) we arrive at
\ba\label{B13}
\{T^K,\check C_M\}=\delta^K_M + \lambda_M^{K L_1 \cdots L_r}\hat C_{L_1}\cdots \hat C_{L_r}+ (r+1) \mu_M^{K L_1 \cdots L_{r}}\hat C_{L_1}\cdots \hat C_{L_r}+O(C^{(r+1)})   \q .
\ea
Therefore we have to choose
\ba\label{B14}
\mu_M^{L_1  \cdots L_{r+1}} \simeq -\frac{1}{(r+1)}\lambda_M^{L_1 \cdots L_{(r+1)}}\simeq -\frac{1}{(r+1)!}\{T^{L_1},\{T^{L_2},\{\cdots,\{T^{L_r},\{T^{L_{r+1}},\hat C_M\}\cdots\}
\ea
in order to satisfy equation (\ref{B10}).

It remains to show that the new constraints commute up to $O(C^{(r+2)})$--terms. To this end consider the Jacobi identity
\ba\label{B15}
\{\{T^K,\check C_M\},\check C_N\}-\{\{T^K,\check C_N\},\check C_M\}=\{\{\check C_N, \check C_M \}, T^K\}  \q .
\ea
Because of property (\ref{B10}) the left hand side is of order $O(C^{(r+1)})$. Using the definition (\ref{B9})) for the $\check C_K$ in terms of the $\hat C_K$ and assumption (\ref{B8a}) for the Poisson bracket between the constraints $\hat C_K$ the right hand side can be written as
\ba\label{B16}
\{\{\check C_N, \check C_M \}, T^K\} &=&\{g_{NM}^{L_1 \cdots L_{r+1}} \check C_{L_1} \cdots \check C_{L_{r+1}},\, T^K\} +O(C^{(r+1)})
\nn \\
&=&(r+1)\, g_{NM}^{L_1 \cdots L_r K} \check C_{L_1} \cdots \check C_{L_{r}}+O(C^{(r+1)})  \q
\ea
where $g_{NM}^{L_1 \cdots L_{r+1}}$ is some set of phase space functions symmetric in the $\{L_1 \cdots L_{r+1}\}$--indices.

Reinserting this result into equation (\ref{B15}) we see that all terms are of the order $O(C^{(r+1)})$ except for $(r+1)\, g_{NM}^{L_1 \cdots L_r K} \check C_{L_1} \cdots \check C_{L_{r}}$. Applying $r$ times the Poisson bracket with the clock variables $T^{K_j}$ with both sides of the equation, we conclude that the $g_{NM}^{L_1 \cdots L_{r+1} }$ have to vanish weakly. This shows that the Poisson bracket between the constraints $\check C_K$ is of order $O(C^{(r+2)})$.
Hence we are able to define a set of constraints that have vanishing Poisson brackets up to terms of arbitrary high order in the constraints. If the iteration procedure converges it will result in a set of Abelian constraints. 

\vspace{1cm}

This method is applicable to an arbitrary choice of Abelian clock variables. Here we are interested in the case where the clock variables are canonically conjugated to the linearized constraints. Let $\{\hat C_K\}_{K=1}^m$ be a constraint set such that 
\ba\label{B17}
\{T^K, \hat C_M\}=\delta^K_M + O(C^r)  \q .
\ea
Note that during the procedure described above to find such constraints the first order terms of the constraints do not change, i.e. ${}^{(1)}\hat C_M={}^{(1)} C_j \delta^j_M$. We have that
\ba\label{B18}
\{T^K, \hat C_M-{}^{(1)}\hat C_M \}=O(C^r)
\ea
is at least of order $r$ in the fluctuations. Therefore if we would rewrite the constraints $\hat C_M $ using the canonical variables $(T^K,{}^{(1)}C_j, Q^d, P_e)$ introduced above, all the terms of order higher than one and lower than $(r+1)$ are independent of the momenta $\{{}^{(1)}C_j\}_{j=1}^m$. Also it is easy to see, that the terms of order lower than $(r+1)$ do not change if one iterates the above procedure to get a set of constraints $\{\hat C_M'\}_{M=1}^m$ satisfying $\{T^K, \hat C_M'\}=\delta^K_M+O(C^s)$ with $s>r$. This again allows us to calculate the completely Abelianized constraints order by order.

If the procedure converges for $r \rightarrow \infty$ we will end up with a deparametrized form of the constraints, i.e. we have constraints ${}^{\infty}C_K$ with
\ba\label{B19}
{}^{\infty}C_K={}^{(1)}C_j\delta^j_K+ E_K( T^M, Q^d,P_e)   \q 
\ea
where the phase space functions $E_K$ are independent from the momenta ${}^{(1)}C_j$. One can show \cite{henneaux}, that constraints of such a form are Abelian. The Abelianization procedure introduced here allows one to compute the deparametrized form of the constraints order by order and is also applicable if the clock variables are quite complicated functions.

In the following we will assume that the Abelianization procedure indeed converges in a finite phase space region around $X_0$ and we will denote by $\check C_K$ the Abelianized constraints. Because of the considerations above we need only finitely many steps to calculate a finite order of $\check C_K$.


\section{Approximate Dirac Observables} \label{appdirac}

Property (\ref{B6}) of the constraints $\tilde C_K$ enables us to construct Dirac observables $F_{[f;T]}(\tau)$ for a set of parameters $\{\tau^K\}_{K=1}^m$ via a (formal) power series: 
\ba\label{B20}
F_{[f;T]}(\tau)= \sum_{r=0}^\infty \frac{1}{r!} \{\cdots\{f, \tilde C_{K_1}\},\cdots,\},\tilde C_{K_r}\}\, (\tau^{K_1}-T^{K_1})\cdots (\tau^{K_r}-T^{K_r})  \q .
\ea
It is straightforward to check that the Poisson bracket of this Dirac observable with the constraints $\tilde C_K$ vanishes at least weakly. The constraints $\tilde C_K$ and $\check C_K$ generate the same flow on the constraint hypersurface, therefore one can replace any of the constraints $\tilde C_M$ in (\ref{B20}) with the constraints $\check C_M$.

The observable $F_{[f;T]}(\tau)$ restricted to the hypersurface $\{T^K=\tau^K\}_{K=1}^m$ is equal to the function $f$. This provides the interpretation for $F_{[f;T]}(\tau)$: Since it is gauge invariant, one can see it as a gauge invariant extension of $f$. Hence $F_{[f;T]}(\tau)$ and  $F_{[f';T]}(\tau)$ coincide at least weakly if $f$ and $f'$ coincide on the hypersurface $\{T^K=\tau^K\}_{K=1}^m$.

If the constraints generate also time evolution (i.e. if time evolution is a gauge transformation) $F_{[f;T]}(\tau,x)$ is known as the complete observable associated to the partial observable (and the clock variables $T^K$). $F_{[f;T]}(\tau,x)$ gives the value of the partial observable $f$ at the ``moment'' at which the clock variables $T^K$ assume the values $\tau^K$.      

Neglecting in the power series (\ref{B20}) all terms of order higher than $q$ in the differences $(\tau^K-T^K)$, i.e. truncating the series to its first $(q+1)$ summands, we obtain a phase space function that has vanishing Poisson brackets with the constraints modulo terms of order $q$ in $(\tau^K-T^K)$ and modulo constraints. In this sense we can obtain approximate Dirac observables, the approximation being good near the gauge fixing hypersurface $\{T^K=\tau^K\}_{K=1}^m$. However in many situations it may be quite involved to find the inverse of the matrix $A^K_j$, which is necessary in order to obtain the new constraints $\tilde C_K$ or $\check C_K$.   

We can also choose to obtain an approximation around a certain phase space point $X_0$ by expanding the complete observable in the fluctuation variables introduced above and by using the specific $\tau^K$ parameters given by $\tau^K=T^K(X_0)$. As already explained this allows us to find the new constraints $\tilde C_K$ or $\check C_K$ order by order since we can invert the matrix $A^K_j$ order by order.\footnote{This applies also to a more general choice of clock variables than the above one, where the clock variables are conjugated to the linearized constraints. A sufficient condition is that the zeroth order of the matrix $A^K_j$ should be invertible.} Denote by ${}^{[q]}F_{[f;T]}(\tau \equiv T(X_0),x)$ the complete observable $ F_{[f;T]}(\tau \equiv T(X_0),x)$ with terms of order higher than $q$ in the fluctuation variables neglected. For the calculation of ${}^{[q]}F_{[f;T]}(\tau \equiv T(X_0),x)$ we will need at most the first $(q+1)$ terms in the series (\ref{B20}) and the constraints $\tilde C_K$ to the $q$--th order.  

The truncated complete observable ${}^{[q]}F_{[f;T]}(\tau \equiv T(X_0))$  will commute with the constraints modulo terms of order $q$ in the fluctuations (and modulo terms vanishing on the constraint hypersurface). Hence we can call ${}^{[q]}F_{[f;T]}(\tau \equiv T(X_0),x)$ an approximate Dirac observable.  

If the power series (\ref{B20}) for the complete observable converges it defines an exact Dirac observable which coincides with the approximate Dirac observable ${}^{[q]}F_{[f;T]}(\tau \equiv T(X_0))$ modulo terms of order $(q+1)$. If the power series does not converge in some phase space region, this will be due to the fact that the clock variables do not provide a good parametrization of the gauge orbits in this phase space region \cite{bd1}. In this case one can try to find a set of new clock variables ${T'}^K$ with a better behaviour in this respect. Assume that the complete observable  $F_{[f';T']}(\tau' \equiv  T'(X_0))$ associated to these new clock variables and the partial observable $f':={}^{[q]}F_{[f;T]}(\tau \equiv T(X_0))$ can be defined. This complete observable will also coincide with ${}^{[q]}F_{[f;T]}(\tau \equiv T(X_0))$ modulo terms of order $(q+1)$, as can be seen by examining the power series (\ref{B20}) for a complete observable and using that $f'$ Poisson commutes modulo terms of order $q$ with the constraints.

\section{Application to General Relativity}   \label{gr}

In this section we will collect all definitions in order to be able to calculate approximate Dirac observables for General Relativity. We will work with the (complex) Ashtekar variables \cite{ashtekarbook} (and use the conventions in \cite{MQG}), because the constraints are polynomial in these variables. However the formalism is independent from the choice of variables.

The canonical variables are fields on a spatial manifold $\Sigma$ the coordinates of which we will denote by $\{\sigma^a\}_{a=1}^3$. We will assume that $\Sigma$ is diffeomorphic to $\Rl^3$.
 The configuration variables are given by a complex connection $\{A^j_a\}_{j,a=1}^3$ where latin letters from the beginning of the alphabet denote spatial indices and from the middle of the alphabet $su(2)$--algebra indices:
\ba\label{E1}
A^j_a=\Gamma^j_a+\beta K^j_a
\ea
where $\beta=i/2$ is the Immirzi parameter, $\Gamma^j_a$ is the spin connection for the triads $e^j_a$ and $K^j_a$ is the extrinsic curvature. The spatial metric can be calculated from the triads by $q_{ab}=e^j_a e^k_b\delta_{jk}$. The conjugated momenta  $E^a_j$ are constructed out of the triads
\ba\label{E2}
E^a_j=\beta^{-1} \epsilon^{a a_1 a_2}\epsilon_{jj_1j_2}e^{j_1}_{a_1}e^{j_2}_{a_2}
\ea
where $\epsilon^{a a_1 a_2}$ and $\epsilon_{jj_1j_2}$ are totally anti--symmetric tensors with $\epsilon_{123}=\epsilon^{123}=1$. The Poisson brackets between these phase space variables are given by
\ba\label{E3}
\{A^j_a(\sigma),E_k^b(\sigma')\}=\kappa \delta^j_k \delta_a^b \delta(\sigma,\sigma')
\ea
where $\kappa=8\pi G_N/c^3$ is the gravitational coupling constant. 

The constraints are given by the Gauss constraints $G_j(\sigma)$, the vector constraints $V_a(\sigma)$ and the scalar constraints $C(\sigma)$:
\ba\label{E4}
G_j &=& \kappa^{-1}(\partial_a E^a_j+\epsilon_{jkl}A^k_aE^a_l) \nn\\
V_a  &=&  \kappa^{-1}F^j_{ab}E^b_j=\kappa^{-1}(\partial_a A_b^j-\partial_bA^j_a +\epsilon_{jkl}A_a^kA_b^l)E^b_j\nn\\
C &=&  \kappa^{-1} \beta^2 F^j_{ab}\epsilon_{jkl} E^a_k E^b_l  \q .
\ea
Note that we use the scalar constraint with density weight two here.

In this work we will choose as our background phase space point Minkowski space $X_0:=({{\bf{A}}_{a}}^j,{{\bf{E}}^b}_k)$ with
\ba\label{A2}
{{\bf{A}}_{a}}^j=0 \q ,\q {{\bf{E}}^b}_k=\beta^{-1}\delta^b_k   \q .
\ea
We will denote the fluctuations around this background by lower case letters:
\ba\label{A3}
{a_a}^j={A_a}^j-{{\bf{A}}_a}^j \q ,\q {e^b}_k={E^b}_k-{{\bf{E}}^b}_k \q .
\ea
Contracting these fluctuation variables with the background triad or its inverse, we can convert the internal Lie algebra indices to spatial indices:
\ba\label{A4}
{a_a}^b:={a_a}^j {{\bf{E}}^b}_k \q ,\q {e^c}_d={e^c}_k {{\bf{E}}^k}_d  \q .
\ea
Since the background metric is flat we can freely raise or lower the spatial indices.
The constraints in these variables are given by
\ba\label{A5}
G_b:=G_j {{\bf E}_b}^j &=& \kappa^{-1}( \partial_a {e^a}_b +\beta \epsilon_{bde} a^{ed} + \beta \epsilon_{bdf} e^{af} {a_a}^d  ) \\
V_a &=&\kappa^{-1}(  \partial_a {a_b}^b-\partial_b {a_a}^b  
  + \beta \epsilon_{fde}{a_a}^d a^{fe} 
+(\partial_a {a_b}^c-\partial_b{a_a}^c){e^b}_c 
+ \beta \epsilon_{fde}{a_a}^d{a_b}^e e^{bf}         ) \q\q \\
 C &=&\kappa^{-1}\beta (  \epsilon^{def}(\partial_a a_{bd}-\partial_ba_{ad})
(\delta^a_e\delta^b_f +2\delta^b_f {e^a}_e +{e^a}_e {e^b}_f)  \nn\\
&& \q\;\;\; + \beta({a_a}^e {a_b}^f-{a_a}^f {a_b}^e )
(\delta^a_e\delta^b_f +2\delta^b_f {e^a}_e +{e^a}_e {e^b}_f)    ) \q .
\ea

The first order of these constraints are the linearized constraints of General Relativity. Linearized constraints arising from a first class constraint set are Abelian \cite{lee}, this applies also to our constraints.

Now we have to choose our clock variables. Having in mind that we have to invert the matrix ${A^K}_j(\sigma,\sigma'):=\{T^K(\sigma),C_j(\sigma')\}$ order by order, it would be convenient (although not necessary) if the zeroth order of this matrix would be given by the identity matrix; i.e. if the clocks would be exactly conjugated to the linearized constraints.
 
 In metric variables such clock variables were already given in the seminal paper \cite{ADM} and used in \cite{kk1} to construct the lowest order ground wave function for quantum gravity. Hence we will call these clock variables ADM clocks. Their interpretation \cite{kk1} is the following: In the gauge where these clock variables vanish (or are constant) the metric is in a coordinate system which is as near to the Cartesian one as possible. Non--constant clock variables mean that the coordinate system is deformed from the optimal one.

These clock variables can be transformed from the metric variables to the Ashtekar variables. This will result in functions which are first order and higher in the fluctuations -- we will keep only the first order terms. Additionally one has to construct a set of clock variables conjugated to the linearized Gauss constraints:
\ba\label{A6}
{}^G T^a &=&
\beta^{-1} {\epsilon^{age}} (-W^{-2} \partial_g \partial_d) {e^d}_e 
+W^{-2} (-\frac{1}{2} W^{-2} \partial^a \partial^d \partial^e + \frac{1}{2} \partial^a \delta^{de}) a_{de}\\
{}^V T^a &=&
-W^{-2}(-W^{-2} \partial^a \partial_d \partial_e +\frac{1}{2} \partial^a \delta_{de}-\partial_e\delta^a_e-\partial_d \delta^a_e)e^{de} \\
{}^C T &=& (4\beta)^{-1} W^{-2}( -\partial^c \epsilon_{cde} e^{de} -\beta (\delta^{de}-W^{-2}\partial^d \partial^e)a_{de})
\ea
where $W:=\sqrt{-\partial_a \partial^a}$. Together with the linearized constraints these clock variables form a set of canonically conjugated variables, i.e. 
$\{{}^GT^a(\sigma), {}^{(1)}G_b(\sigma')\}=\delta^a_b \delta(\sigma, \sigma')$ and so on. Here the superscript ${}^{(1)}$ to the left of a symbol denotes its first order. The Poisson brackets between two clock variables vanish. 

A tensor mode decomposition\footnote{For the definition of tensor modes see appendix \ref{tensor modes}.} of the linearized constraints and the clock variables reveals that these include all tensor modes except for the symmetric trace--free transverse (STT) modes, which are the physical degrees of freedom for linearized gravity:
\ba\label{A7} 
{}^{(1)}G_b &=& \kappa^{-1}\big(\delta _{bc} \partial_a ({}^{LT}P^{ac}_{de}+{}^{LL}P^{ac}_{de})e^{de}+\beta \epsilon_{bcd}( {}^{LT}P^{dc}_{af}+{}^{TL}P^{dc}_{af}+{}^{AT}P^{dc}_{af}       )a^{af}  \big) \nn\\
{}^{(1)}V_a &=& \kappa^{-1}\big( \delta^{bc}(\partial_a {}^{T}P^{de}_{bc}-\partial_b {}^{TL}P^{de}_{ac})a_{de}                \big)  \nn\\
{}^{(1)}C &=&\kappa^{-1}\big(2\beta \epsilon^{abd} \partial_a {}^{AT}P^{fe}_{bd}a_{fe}  \big)
\nn\\ 
{}^GT^a &=&
\beta^{-1} \delta^{af}\epsilon_{fbc} {}^{LT}P^{bc}_{de} e^{de}+ W^{-2}(\delta^{fa}\partial^c\, {}^{LL}P^{de}_{cf}+\tfrac{1}{2}\delta^{cb}\partial^a \,{}^T P^{de}_{bc} )a_{de} \nn\\
{}^VT^a &=& 
-W^{-2}(-\partial_b\, {}^{LT}P^{ba}_{de}-\partial_b\, {}^{TL}P^{ab}_{de}+\tfrac{1}{2}\partial^a\, {}^{T}P^{cf}_{de}\delta_{cf}-\tfrac{1}{2}\partial_b {}^{LL}P^{ba}_{de}) e^{de}
\nn\\
{}^{C}T &=&(4\beta)^{-1}W^{-2}(-\partial^c \epsilon_{cab} {}^{AT}P^{ab}_{de}e^{de}+\beta (-\delta^{cb}\,{}^{T}P^{de}_{cb}-2 \delta^{cb}\,{}^{LL}P^{de}_{cb} )\,a_{de})  \q .
\ea
Indeed one can introduce a new canonical coordinate system, with coordinates given by the $STT$--modes of $a_{ab}$ and $e^{cd}$, the linearized constraints and the clock variables. We will call these variables $STT$--modes, $C$--modes and $T$--modes respectively.

\subsection{Asymptotic conditions} \label{asymp}

Since we want to work with asymptotically flat spacetimes we have to formulate asymptotic conditions for our phase space variables. These asymptotic conditions have to ensure \cite{beig,thiemanna} that the symplectic structure is well defined. One such choice is to impose the conditions \cite{thiemanna}
\ba\label{f1}
a_{ab} &\rightarrow& \frac{B_{ab}}{r^2}+O(r^{-3}) \nn\\
e^{ab} &\rightarrow& \frac{F^{ab}}{r^1}+O(r^{-2})
\ea
where $r$ is an asymptotic spherical coordinate, $B_{ab}$ is a smooth tensor on spatial infinity with odd parity and $F^{ab}$ a smooth tensor on spatial infinity with even parity. 

With this choice of boundary conditions it is well defined to integrate the Gauss constraint with a smearing function $\Lambda^j$ which is $O(r^{-2})$ and the vector and scalar constraints with smearing functions which are $O(r^{-1})$. However we will also have to deal with smearing functions which are $O(r^{-1})$ and have even parity in leading order for the Gauss constraint and are $O(r^0)$ with odd parity in leading order for the vector and scalar constraint respectively. Additionally we will have the case of a constant smearing function for the scalar constraint. Such smearing functions arise if we consider the clock variables as smearing functions. In order to make the constraints with such smearing functions functionally differentiable, i.e. in order to be able to perform integration by parts, we will follow the usual strategy \cite{beig, thiemanna} and subtract from the constraints those boundary terms that arise if one performs integration by parts.

Note that there is only a divergence problem for the lowest order part of the constraints, i.e. the linearized constraints. This holds also for the various new constraints $\tilde C_K$ and $\check C_K$ -- the divergence problem exists only for the linear order part and these parts coincide with the linear order parts of the original constraints. Hence it is sufficient to consider this part of the constraints. Boundary terms which arise from the higher order parts of the constraints vanish due to the asymptotic conditions. The smeared linearized constraints with boundary terms subtracted are
\ba
{}^{(1)}G[\Lambda] &:=&
\kappa^{-1}\int_\Sigma \Lambda^b (\partial_a {e^a}_b +\beta \epsilon_{bde}a^{ed}) \bd \sigma
-\kappa^{-1}\int_{\partial \Sigma} \Lambda^b {e^a}_b \bd S_a \nn \\
&=&
\kappa^{-1}\int_\Sigma (-\partial_a \Lambda^b){e^a}_b +\Lambda^b \beta \epsilon_{bde}a^{ed} \bd \sigma  
\\
{}^{(1)}V[N] &:=&
\kappa^{-1}\int_\Sigma N^a (\partial_a {a_b}^b-\partial_b {a_a}^b) \bd \sigma -\kappa^{-1} \int_{\partial \Sigma}  (N^a {a_b}^b - N^b {a_b}^a )\bd S_a \nn \\
&=& -\kappa^{-1} \int_\Sigma  (( \partial_a N^a)  {a_b}^b-(\partial_b N^a) {a_a}^b) \bd \sigma \\
{}^{(1)}C[N] &:=& 2\kappa^{-1}\beta \int_\Sigma \epsilon^{abc} N \partial_a a_{bc} \bd \sigma -  2\kappa^{-1}\beta  \int_{\partial \Sigma} N \epsilon^{abc}  a_{bc} \bd S_a   \nn \\      
\label{f3}
&=&- 2\kappa^{-1}\beta \int_\Sigma \epsilon^{abc} (\partial_a N)a_{bc} \bd \sigma 
\ea
where $\bd S_a=n_a \bd S$, $\bd S$ is the volume element of the sphere at infinity and $n_a$ is the outward pointing unit normal and has odd parity. Hence the boundary terms actually vanish for $\Lambda^b$ even and $N,N^a$ odd. (Here it is understood that the integration is performed over some coordinate ball with finite radius and that one then considers the limit $r \rightarrow \infty$).

The linearized part of $C[N]$ vanishes for a constant $N\equiv 1$ smearing function. Also we have that for such a smearing function the boundary term $\int_{\partial \Sigma}  \epsilon^{abc}  a_{bc} \bd S_a$ is equal to the ADM energy modulo terms that are proportional to the Gauss constraints \cite{ADM,ashtekarbook,thiemanna} and therefore it is a Dirac observable. On the constraint hypersurface this ADM energy is equal to ${}^{(2+)}C[1]=\int_\Sigma {}^{(2+)}C \bd \sigma$.

Henceforth we will understand that we will use the scalar constraints with boundary terms as in (\ref{f3}). Hence we can perform integration by parts if we have a smearing function approaching a constant for $r \rightarrow \infty$. For smearing functions with stronger fall-off at infinity or for $\Lambda^b$ even parity and $O(r^{-1})$, $N,N^a$ odd parity  and $O(r^0)$ the boundary terms vanish.

\section{Dynamics} \label{dyn}

The clock variables and the expanded constraints from the last section allow us to compute perturbative complete observables order by order. First we have to determine the constraints 
\ba\label{d1}
\tilde C_K(\sigma):=\int_\Sigma \bd \sigma'  C_j(\sigma') (A^{-1})^j_K(\sigma',\sigma) \q .
\ea
Here the index $j$ runs from $0$ to $6$, $C_0:=C$, $C_1$ to $C_3$ is equal to the vector constraints and $C_4$ to $C_6$ to the Gauss constraints. The index $K$ runs also from $0$ to $6$ with $T^0:={}^{C}T$, and so on. 

For (\ref{d1}) we need the inverse of the matrix
\ba\label{A8}
A^K_j(\sigma,\sigma'):=\{T^K(\sigma),C_j(\sigma')\}&=& \delta^K_j\delta(\sigma,\sigma')+\{T^K(\sigma), {}^{(2+)}C_j(\sigma')\} \nn \\
&=:&\delta^K_j\delta(\sigma,\sigma')+B^K_j(\sigma,\sigma')
\ea
where ${}^{(2+)}C_j(\sigma')$ denotes all terms of order two or higher of the constraint $C_j(\sigma')$. Hence $A^K_j(\sigma,\sigma')$ is to lowest order given by the (infinite dimensional) identity matrix. This allows us to invert the matrix order by order using an iterative equation:
\ba\label{A9}
(A^{-1})^l_M(\sigma,\sigma')=                                                                       \delta^l_M(\sigma,\sigma') -    \int_\Sigma \left( \delta^l_K B^K_j(\sigma,\sigma'') (A^{-1})^j_M(\sigma'',\sigma')  \right)  \bd \sigma''    \q .
\ea

Hence we can determine the constraints $\tilde C_K$ up to a finite order in the fluctuations in a finite number of steps. To calculate the constraints $\check C_K$ defined in section \ref{abel} to some order $r$ we have to perform $(r-1)$ times the iteration step described in section \ref{abel}. For instance the second order of the constraints $\check C_K$ is given by
\ba\label{d2}
{}^{(2)}\check C_K(\sigma)={}^{(2)}\tilde C_K(\sigma)+\int_\Sigma \bd \sigma' \int_\Sigma \bd \sigma'' \,\, {}^{(0)}\mu_K^{L_1 L_2}(\sigma,\sigma',\sigma'') {}^{(1)}\tilde C_{L_1}(\sigma')\, {}^{(1)}\tilde C_{L_2}(\sigma'')
\ea
with
\ba\label{d3}
{}^{(0)}\mu_K^{L_1 L_2}(\sigma,\sigma',\sigma'')=-\frac{1}{2}\{T^{L_1}(\sigma'),\{T^{L_2}(\sigma''),{}^{(2)}\tilde C_K(\sigma)\}\} \q .
\ea

 This can be used to calculate the complete observable $F_{[f;T]}$ with parameter choices $\tau^K \equiv 0$ to some arbitrary finite order $m$.
\ba\label{d4}
{}^{[m]}F_{[f;T]}(\tau^K\equiv 0)&=&\sum_{r=0}^m \frac{1}{r!} \int_\Sigma \bd \sigma_1 \cdots \bd \sigma_r \{\cdots\{f, {}^{[m]}\check C_{K_1}(\sigma_1)\},\cdots\},{}^{[m-r+1]}\check C_{K_r}(\sigma_r) \} \times 
\nn \\ 
&&\q\q\q\q\q\q\q\q\q\q(-1)^r T^{K_1}(\sigma_1)\cdots T^{K_r}(\sigma_r) \q .
\ea

However we are also interested in dynamical questions, that is complete observables for varying clock parameters $\tau$. Introducing non--vanishing clock parameters into the series for the complete observable (\ref{B20}) we see that it is now a power series in $(\tau^K-T^K)$ which includes the zeroth order term $\tau^K$. Hence the complete observable to the $m$--th order is not a finite sum anymore.

Let us consider the power series for complete observables for non--vanishing clock parameters $\tau$ in more detail, separating terms with different powers of $T^K$:
\ba\label{A10}
F_{[f;T]}(\tau) &\simeq &\sum_{r=0}^\infty \frac{1}{r!}  \int_\Sigma \bd\sigma_1 \cdots \bd\sigma_r \{\cdots\{f, \check C_{K_1}(\sigma_1)\},\cdots,\check C_{K_r}(\sigma_r)\} \times \nn\\
&& \q\q (\tau^{K_1}(\sigma_1)-T^{K_1}(\sigma_1))\cdots (\tau^{K_r}(\sigma_r)-T^{K_r}(\sigma_r)) 
\nn\\
&\simeq& \sum_{r=0}^\infty \frac{1}{r!}  
 \int_\Sigma \bd\sigma_1 \cdots \bd\sigma_r 
\{\cdots\{f, \check C_{K_1}(\sigma_1)\},\cdots,\check C_{K_r}(\sigma_r)\} \times \nn\\
&& \q\q
 \tau^{K_1}(\sigma_1) \tau^{K_2}(\sigma_2) \cdots \tau^{K_r}(\sigma_r) \, +
\nn \\
&&
  \sum_{r=1}^\infty \frac{1}{r!}
 \int_\Sigma \bd\sigma_1 \cdots \bd\sigma_r 
\{\cdots\{f, \check C_{K_1}(\sigma_1)\},\cdots,\check C_{K_r}(\sigma_r)\}) \times \nn\\
&& \q\q
\sum_{q=1}^r \tau^{K_1}(\sigma_1) \tau^{K_2}(\sigma_2) \cdots  (-T^{K_q}(\sigma_q))\cdots\tau^{K_r}(\sigma_r) \, +
\nn \\
&&  \sum_{r=2}^\infty \frac{1}{r!} 
\int_\Sigma \bd\sigma_1 \cdots \bd\sigma_r 
\{\cdots\{f, \check C_{K_1}(\sigma_1)\},\cdots,\check C_{K_r}(\sigma_r)\} \times \nn\\
&& \q\q
 \sum_{q=1,p=2,q<p}^r \tau^{K_1}(\sigma_1) \tau^{K_2}(\sigma_2) \cdots  (-T^{K_q}(\sigma_q))\cdots   (-T^{K_p}(\sigma_p))\cdots\tau^{K_r}(\sigma_r)
\nn\\
&& +\ldots
\ea
Since the constraints $\check C_{K_i}(\sigma)$ Poisson commute up to terms at least quadratic in the constraints among themselves, one can rearrange the constraints $\check C_{K_i}(\sigma)$ in the formula above in an arbitrary order, changing the expression only by terms proportional to the constraints. This allows us to write (\ref{A10}) in two ways. On the one hand 
\ba\label{A11}
F_{[f;T]}(\tau^K) &\simeq&
 \sum_{r=0}^\infty \frac{1}{r!} \int_\Sigma \bd\sigma_1 \cdots \bd\sigma_r 
 \{\cdots\{f, \check C_{K_1}(\sigma_1)\},\cdots,\check C_{K_r}(\sigma_r)\} \times \nn\\
&& \q\q
\tau^{K_1}(\sigma_1) \tau^{K_2}(\sigma_2) \cdots \tau^{K_r}(\sigma_r)\, +
\nn \\
&& \sum_{r=1}^\infty \frac{1}{(r-1)!} \int_\Sigma \bd\sigma_1 \cdots \bd\sigma_r 
\{\cdots\{f, \check C_{K_1}(\sigma_1)\},\cdots,\check C_{K_r}(\sigma_r)\}(x) \times \nn\\
&&\q\q
  \tau^{K_1}(\sigma_1) \tau^{K_2}(\sigma_2) \cdots \cdots\tau^{K_{r-1}}(\sigma_{r-1})(-T^{K_r}(\sigma_r))\,+
\nn \\
&& \sum_{r=2}^\infty \frac{1}{2!(r-2)!} \int_\Sigma \bd\sigma_1 \cdots \bd\sigma_r 
\{\cdots\{f, \check C_{K_1}(\sigma_1)\},\cdots,\check C_{K_r}(\sigma_r)\}(x) \times \nn\\
&& \q\q
 \tau^{K_1}(\sigma_1) \tau^{K_2}(\sigma_2) \cdots  \tau^{K_{r-2}}(\sigma_{r-2})       (-T^{K_{r-1}}(\sigma_{r-1}))  (-T^{K_r}(\sigma_r))
\nn\\
&& +\ldots
\ea
which after a relabelling of the summation index $r$ can be recognized as 
\ba\label{A12}
F_{[f;T]}(\tau^K) \simeq F_{[ \alpha_{\check C_K}^{\tau^K}(f);T]}(\tau^K \equiv 0)  \q .
\ea
Here $\alpha_{\check C_K}^{\tau^K}(f)$ denotes the evolution of $f$ with respect to the constraints $\check C_K(\sigma)$ and the parameters $\tau^K(\sigma)$, given by the first summand in (\ref{A11}). 

In the same way one can arrive at
\ba\label{A13}
F_{[f;T]}(\tau,x) \simeq \sum_{q=0}^\infty \frac{1}{q!} \int_\Sigma \bd\sigma_1 \cdots \bd\sigma_q && \alpha^{\tau^M}_{\check C_M}\big(\{\cdots\{f, \check C_{K_1}(\sigma_1)\},\cdots,\check C_{K_q}(\sigma_q)\}\big)(x) \times \nn\\
&& (-T^{K_1}(\sigma_1)) (-T^{K_2}(\sigma_2)) \cdots (-T^{K_q}(\sigma_q))  \q .
\ea

Consider the complete observable in the form (\ref{A12}). There one has first to evolve the phase space function $f$ with respect to the constraints $\check C_K(\sigma)$ and the parameters $\tau^K(\sigma)$ and afterwards to calculate the complete observable. If one is interested in an $m$-th order approximation, then the difficulty arises that in general the $m$-th order $\alpha_{\check C_K}^{\tau^K}(f)$ will contain infinitely many terms involving arbitrarily high order of the constraints. The reason for this is, that  the lowest order in the Poisson bracket $\{{}^{(m)}g,\check C_K(\sigma)\}$ of an $m$-th order function ${}^{(m)}g$ with the constraints is $(m-1)$, hence one looses one order due to the first order part of the constraints.

However we are not interested in the complete observable for arbitrary parameter values $\tau^K(\sigma)$, for dynamical questions it is sufficient to be able to calculate complete observables for a one-parameter family of parameters. For our case a natural choice is $\tau^0(t;\sigma)\equiv t$ and $\tau^K(t;\sigma)\equiv 0$ for all $K \neq 0$. The parameter $t\in \Rl$ would correspond to a notion of time that is as near as possible to the Minkowskian time of the background and as we will see later on corresponds to time translation at infinity. For this choice of parameters the evolution $\alpha_{\check C_K}^{\tau^K}(f)$ becomes
\ba\label{A14}
\alpha_{\check C_0}^t(f):=\alpha_{\check C_K}^{\tau^K(t)}(f)=\sum_{r=0}^\infty \frac{t^r}{r!}\{\cdots\{f,\check C_0[1]\},\cdots, \check C_0[1]\} 
\ea
where $\check C_0[1]:=\int_\Sigma \check C(\sigma) \, \bd \sigma -  2\kappa^{-1}\beta  \int_{\partial \Sigma} \epsilon^{abc}  a_{bc} \bd S_a $ according to the definition in section \ref{asymp}. Now the first order part of $\check C_0[1]$ is equal  to the first order part of $C[1]$ and hence vanishes
according to equation (\ref{f3}).

Therefore we are left with an evolution of $f$ with respect to a generating function ${}^{(2+)}\check C[1]$ which is at least second order. This allows us to compute the $m$--th order approximation to this evolution in a systematic manner -- the highest order term required of ${}^{(2+)}\check C[1]$ is the $(m+1)$--th order term (for $f$ a first order quantity).

In order to calculate the $m$-th order approximation of the complete observable (\ref{A12}) it is sufficient to have the $m-th$ order approximation of $\alpha_{\check C_0}^t(f)$: The complete observable (for parameters $\tau^K$ set to zero) of an $n$--th order quantity is at least of $n$--th order, that is one does not loose any orders in the second step of the calculation in (\ref{A12}). Moreover the calculation of the $m$--th order complete observable (again for parameters $\tau^K$ set to zero) requires at most $(m+1)$ terms in the power series for complete observables. 

Before considering explicitly the second order approximation of the complete observable (\ref{A12}) we will remark on the formula (\ref{A13})
\ba\label{A16}
F_{[f;T]}(\tau^K) \simeq \sum_{q=0}^\infty \frac{1}{q!} \int_\Sigma \bd\sigma_1 \cdots \bd\sigma_q && \alpha^{\tau^M}_{\check C_M}\big(\{\cdots\{f, \check C_{K_1}(\sigma_1)\},\cdots,\check C_{K_q}(\sigma_q)\}\big) \times \nn\\
&& (-T^{K_1}(\sigma_1)) (-T^{K_2}(\sigma_2)) \cdots (-T^{K_q}(\sigma_q))  
\ea
 of the complete observable and bring it into a form where the ADM energy appears explicitly as the generator for the evolution of the complete observables in the time parameter $t$, that is as the physical Hamiltonian \cite{t1}. To this end choose also here the one--parameter family $\tau^0(t,\sigma)\equiv t$ and $\tau^K(t,\sigma)\equiv 0$ for $K\neq 0$. The first order term ${}^{(1)}\check C_0[1]$ vanishes, that is we can replace $\alpha^{\tau^M}_{\check C_M}$ in (\ref{A16}) by $\alpha^t_{{}^{(2+)}\check C_0}$. Now the functions ${}^{(2+)}\check C_0[1]$ Poisson commute with the clock variables $T^K(\sigma)$ since ${}^{(2+)}\check C_0(\sigma)$ does not contain any $C$--modes, i.e. does not depend on the variables canonically conjugated to the clock variables.

Hence $\alpha^t_{{}^{(2+)}\check C_0}(T^K(\sigma))= T^K(\sigma)$ and we can write (\ref{A16}) as
\ba\label{A17}
F_{[f;T]}(t) \simeq \alpha^t_{{}^{(2+)}\check C_0} \bigg(  F_{[f;T]}(t=0)          \bigg) \q ,
\ea
where we use $t$ as an abbreviation for the one--parameter family $\tau^0(t,\sigma)\equiv t$ and $\tau^K(t,\sigma)\equiv 0$ for $K\neq 0$. In equation (\ref{A17}) we evolve the Dirac observable $F_{[f;T]}(t=0)$ with respect to the generating function ${}^{(2+)}\check C_0[1]$. However since we evolve a Dirac observable we can add any combination of constraints to the generating function, that is we could also use ${}^{(2+)}C[1]$ which differs from ${}^{(2+)}\check C_0[1]$ only by terms proportional to the constraints.

Now ${}^{(2+)}C[1]$ (and ${}^{(2+)}\check C_0[1]$) coincides on the constraint hypersurface with the ADM energy. This shows that the ADM energy appears as the physical Hamiltonian, that is as the generating function for the time evolution chosen here. Since the ADM energy generates time translations at infinity, we see that our choice of time parameter corresponds to time translations at infinity.

\section{ The second order approximation}  \label{approximation}

In this section we will consider explicitly the second order approximation of the complete observable
\ba\label{A19}
F_{[f;T]}(t) \simeq F_{[ \alpha_{{}^{(2+)}\check C_0}^{t}(f);T]}(0)  \q 
\ea
where $f$ is a first order phase space function and we will assume that it commutes with the linearized constraints and with the clock variables. For pure gravity we could choose $f= {}^{STT}P^{ab}_{cd}a_{ab}(\sigma)=:{}^{STT}a_{cd}(\sigma) $ or $f= {}^{STT}P^{ab}_{cd}e_{ab}(\sigma)=:{}^{STT}e_{cd}(\sigma) $. The higher than second order approximations can be obtained in a similar manner. To simplify notation we will introduce $\check H:={}^{(2+)}\check C_0[1]$. The calculation of (\ref{A19}) proceeds in two steps, first one has to calculate the evolution of $f$ with respect to $\check H$ and afterwards one has to compute the complete observable corresponding to $\alpha_{\check H}^{t}(f)$ with the clock parameters $\tau^K$ set to zero. The complete observable $F_{[g;T]}(0,x)$ of a phase space function of order $m$ is at least of order $m$, hence one does not loose any orders in the second step. This means that in order to calculate (\ref{A19}) to second order we also need $\alpha_{\check H}^{t}(f)$ to second order:
\ba\label{A20}
{}^{[2]}\alpha_{\check H}^{t}(f)
&=&
\alpha_{{}^{(2)}\check H}^{t}(f)+ \sum_{r=1}^\infty \frac{t^r}{r!} \sum_{s=0}^r\{\{\{f,{}^{(2)}\check H \}_s,\, {}^{(3)}\check H \}\, {}^{(2)}\check H \}_{r-s-1}  \nn\\
&=&
\alpha_{{}^{(2)}\check H}^{t}(f)+ \sum_{p,q=0}^\infty \frac{t^{(p+q+1)}}{(p+q+1)!}\{\{\{f,{}^{(2)}\check H \}_p,\, {}^{(3)}\check H \}\, {}^{(2)}\check H \}_{q} 
 \q .
\ea
Using the identity
\ba\label{A21}
\frac{t^{(q+p+1)}}{(q+p+1)!}=  \int^t_0  \frac{(t-t')^q}{q!} \frac{{t'}^p}{p!} \bd t'
\ea
for $q,p$ natural numbers, we can rewrite the sum in (\ref{A20}) as
\ba\label{A21a}
{}^{[2]}\alpha_{\check H}^{t}(f)
&=&
\alpha_{{}^{(2)}\check H}^{t}(f)+ \int_0^t \bd t' \sum_{p,q=0}^\infty \frac{(t-t')^q}{q!}\frac{{t'}^p}{p!} \{\{\{f,{}^{(2)}\check H \}_q,\, {}^{(3)}\check H \}\, {}^{(2)}\check H \}_{p}  \nn\\
&=& \alpha_{{}^{(2)}\check H}^{t}(f)+ \int_0^t \bd t' \, \,  \alpha_{{}^{(2)}\check H}^{(t')}  \big( \,\,   \{\alpha_{{}^{(2)}\check H}^{(t-t')}(f), {}^{(3)}\check H\} \, \,\big)  \q .
\ea

For higher order calculations one has to use the identity (\ref{A21}) iteratively. 
Note that $\alpha_{{}^{(2)}\check H}^{t}$ is the propagator for a linear field theory, hence can be given explicitly.


Finally we have to compute the second order complete observable corresponding to ${}^{[2]}\alpha_{\check H}^{t}(f)$. The second order complete observable corresponding to a first order function $g$ has three summands:
\ba\label{A22}
{}^{[2]}F_{[g;T]}(0) 
&=&
 g+ \int_\Sigma \bd \sigma_1 \{g, {}^{[2]}\check C_{K_1}(\sigma_1) \} (-T^{K_1}(\sigma_1)) \nn\\
   && + \frac{1}{2}\int_\Sigma \bd \sigma_1 \,\bd \sigma_2 \{ \{ g, {}^{[2]}\check C_{K_1}(\sigma_1) \}, {}^{[1]}\check C_{K_2}(\sigma_2) \} (-T^{K_1}(\sigma_1)) (-T^{K_2}(\sigma_2))  \q .
\ea

If $g$ is a second order quantity, we only need the first two summands and replace ${}^{[2]}\check C_{K_1}(\sigma_1)$ by ${}^{[1]}\check C_{K_1}(\sigma_1)$ there. Alternatively since we have that (\cite{bd1})
\ba\label{A23}
F_{[g_1 \cdot g_2 ;T]}(0) &=& F_{[g_1 ; T]} (0) \, \cdot \,F_{[g_2 ; T]}(0) \nn\\
F_{[g_1 + g_2 ;T]}(0) &=& F_{[g_1 ; T]} (0)  +F_{[g_2 ; T]}(0)
\ea
we can calculate the second order complete observable of a second order function by computing the first order complete observable of its first order factors.

Note that the complete observables associated to $g$ and $g'$, i.e. $F_{[g;T]}(0,x)$ and $F_{[g';T]}(0,x)$ coincide weakly if $g$ and $g'$ coincide weakly on the (gauge fixing) hypersurface $\{T^K \equiv 0\}$. Hence before we calculate the complete observable associated to $\alpha^t_{\check H}(f)$ we can set all terms vanishing on the gauge fixing hypersurface to zero. Therefore we also need to determine $\alpha^t_{\check H}(f)$ only modulo such terms.

\subsection{The two lowest orders of the Hamiltonian} \label{propa}

Here we will consider the second and third order of the Hamiltonian $\check H$ in more detail. 
We know that any finite order of $\check H:={}^{(2+)}\check C_0[1]$ does not contain any $C$--modes. On the other hand, as explained in the last section, we can omit in $\alpha^t_{\check H}(f)$ any terms containing $T$--modes. We need therefore to consider $\check H$ only up to terms containing $T$--modes. Let us denote by $\check H'$ the function obtained from $\check H$ by setting the $T$--modes to zero. Then we can use $\check H'$ instead of $\check H$ in the propagation of $f$ and furthermore we know that the only gravitational modes appearing in any finite order of $\check H'$ are the $STT$--modes. 

These consideration simplify very much the calculation of at least the lower order terms of $\check H'$ since one can omit all terms which contain either $C$--modes or $T$-modes. This is the advantage provided by introducing the constraints $\check C_K$ as compared to the constraints $\tilde C_K$.

Consider the second order of $\check H$. The second order of $\tilde C_0[1]$ is given by
\ba\label{C1} 
{}^{(2)}\tilde C_0[1]= {}^{(2)} C[1]+ \int_\Sigma \,  {}^{(1)}C_j(\sigma')\, {}^{(1)}(A^{-1})^j_0(\sigma',\sigma)\, \bd \sigma' \bd \sigma  \q .
\ea
For the second order of $\check H=  {}^{(2+)}\check C_0[1]$ we have to add a term quadratic in the first order constraints:
\ba\label{C2}
{}^{(2)}\check C_0[1]=
{}^{(2)}\tilde C_0[1]-\frac{1}{2!} \int \{T^{L_1}(\sigma_1),   \{T^{L_2}(\sigma_2),  {}^{(2)}\tilde C_0(\sigma_3)\}\,  {}^{(1)}\tilde C_{L_1}(\sigma_1) \,  {}^{(1)}\tilde C_{L_2}(\sigma_2) \bd \sigma_1 \bd \sigma_2 \bd \sigma_3 \q .\q
\ea

Hence the second order of $\check H$ is given by the second order of $C[1]$ plus terms which contain $C$--modes. Therefore we can obtain the second order of $\check H'$ by setting all $T$--modes and $C$--modes in the second order of $C[1]$ to zero.  

For the third order we need
\ba\label{C3} 
{}^{(3)}\tilde C_0[1]&=& {}^{(3)} C[1]+ \int_\Sigma \,  {}^{(2)}C_j(\sigma')\, {}^{(1)}(A^{-1})^j_0(\sigma',\sigma)\, \bd \sigma' \bd \sigma + \nn\\
&&
\int_\Sigma \,  {}^{(1)}C_j(\sigma')\, {}^{(2)}(A^{-1})^j_0(\sigma',\sigma)\, \bd \sigma' \bd \sigma
 \q .
\ea
Here the second summand on the right hand side might turn out to be relevant for $\hat H'$. However as shown in appendix \ref{FA} the functions $\int {}^{(1)}(A^{-1})^j_0(\sigma',\sigma) \bd \sigma$ contain only $T$-- and $C$--modes. Furthermore one can check that the terms one has to add to ${}^{(3)}\tilde C_0[1]$ to arrive at the third order of $\check C_0[1]$ are at least linear in the $C$--modes. Therefore we can obtain also the third order of $\check H'$ by restricting ${}^{(3)} C[1]$ to the $STT$--modes. However the fourth order of $\check H'$ will differ from ${}^{(4)} C[1]$ restricted to the $STT$--modes. In particular in this order terms with the inverse derivative operator $W^{-2}=(-\partial_a \partial^a)^{-1}$ will arise, leading to a non--local time generator $\check H'$. This reflects the non--locality of our choice of time function.

\section{Gravity coupled to a scalar field} \label{scalar}

In this section we will consider gravity coupled to a scalar field and compute the complete observable associated to the scalar field to second order. Here the scalar field is assumed to have only small deviations from the zero value, that is the scalar field and its conjugated momentum will be counted as phase space functions of first order. As we will see, the first order complete observable coincides with the expression for the scalar field on a fixed Minkowski background. Hence we will compute the lowest order gravitational correction to this expression. 

In order to couple a scalar matter field to gravity we have to add the following matter contributions to the vector and scalar constraints (\ref{E4}):
\ba\label{m1} 
{}^{\phi}V_b &=&\frac{1}{\gamma} \pi \partial_b \phi  \\
{}^{\phi}C &=& \frac{1}{2\gamma} \left( \pi^2 + q q^{ab} \partial_a \phi \partial_b \phi + q m^2 \phi^2\right) \label{m1a}
\ea
where $\phi$ and $\pi$ denote the scalar field and its conjugated momentum with $\{\phi(\sigma),\pi(\sigma')\}=\gamma \delta(\sigma, \sigma')$. Here $\gamma$ is a coupling constant for the scalar field and $m$ is the mass for the scalar field. We will assume that these fields fall off as $O(r^{-2})$ for $r\rightarrow\infty$. Furthermore $q^{ab}$ is the inverse metric and $q=\text{det}(q_{ab})$ the determinant of the metric. With 
\ba\label{m2}   
 q q^{ab}=\beta^2 E^a_j E^b_j \q\q\q\q\q q=\beta^3 \text{det}(E^a_j)
\ea
the scalar matter constraint (\ref{m1a}) can be expanded to
\ba\label{m3}
{}^{\phi}C &=&\frac{1}{2\gamma} \big( 
    \pi^2 + (\delta^{ab}+2e^{ab}+e^{ad}{e^b}_d)\,\, \partial_a \phi \partial_b \phi + \nn\\
&&\q\q \q     ( 1+{e^b}_b + \frac{1}{2}({e^a}_a {e^b}_b-{e^a}_b {e^b}_a)+ \frac{1}{3!}\epsilon^{abc}\epsilon^{def}e_{ad}e_{be}e_{cf} )\,\, m^2 \phi^2 \big)  \q .
\ea

The lowest order of (\ref{m3}) is the second order and coincides with the Hamiltonian for a scalar field on a flat space--time. The third order has a gravitational correction. We need this third order for the third order of the propagator $\check H'$ as defined in section \ref{propa}. As explained there we just need to restrict the third order of (\ref{m3}) to the $STT$-modes, hence the matter contribution to the third order of $\check H'$ is
\ba\label{m4} 
{}^{(3)}{}^{\phi}\check H'=\frac{1}{\gamma} {}^{STT}e^{ab}\partial_a \phi \partial_b \phi  \q .
\ea

Let us calculate the first and second order of the propagated scalar  field according to (\ref{A21}). The first order is given by
\ba\label{m5}
{}^{(1)}\alpha_{\check H'}^{t}(\phi(\sigma))
&=& \sum_{r=0} \frac{t^r}{r!}\{\phi(\sigma), {}^{(2)}{}^{\phi}C[1]\}_r \nn\\
&=& \int_\Sigma S(t,\sigma;0,\sigma') \pi(\sigma')+S'(t,\sigma;0,\sigma') \phi(\sigma') \, \bd \sigma'
\ea
where we introduced the propagators $S$ and $S'$ for the scalar field. These propagators and the propagators $G,G'$ and $G''$ for the gravitons are reviewed in appendix \ref{propagator}.
For the second order of the propagated scalar field we find
\ba\label{m6}
{}^{(2)}\alpha_{\check H'}^{t}(\phi(\sigma))
&=& 
 \int_0^t \bd t' \, \,  \alpha_{{}^{(2)}\check H'}^{(t'}  \big( \,\,   \{\alpha_{{}^{(2)}\check H'}^{(t-t')}(\phi(\sigma)), {}^{(3)}\check H'\} \, \,\big)  \nn \\
&=&
\int_0^t \bd t' \, \,  \alpha_{{}^{(2)}\check H'}^{t'}  \big( \,\, 
 \int_\Sigma \, 2\, S(t-t',\sigma;0,\sigma')\, ({}^{STT}e^{ab}\,\partial_a \partial_b \phi)(\sigma') \bd \sigma'   
       \, \,\big) \nn \\
&=&
\int_0^t \bd t' \, \,   \int_\Sigma  \bd \sigma'\,2\, S(t-t',\sigma;0,\sigma')\,  \times \nn\\
&&
\int_\Sigma  \big( {G'}^{ab}_{cd}(t',\sigma';0,\sigma'') {}^{STT}a^{cd}(\sigma'')+{G''}^{ab}_{cd}(t',\sigma';0,\sigma'') {}^{STT}e^{cd}(\sigma'') \big) \bd \sigma''  \times \nn \\
&&
\partial_a^{\sigma'}\partial_b^{\sigma'} \int_\Sigma \big( S(t',\sigma';0,\sigma''') \pi(\sigma''')+ S'(t',\sigma';0,\sigma''') \phi(\sigma''')\big) \bd \sigma'''  \q .
\ea

The two contributions (\ref{m5}) and (\ref{m6}) to the propagation of the scalar field can be summarized by introducing new propagator functions for the scalar field that depend on the gravitational variables
\ba\label{m6b}
S_{gr}(t,\sigma;0,\sigma')&:=&S(t,\sigma;0,\sigma')+ 
\nn \\
 && \int_0^t \bd t' \, \,   \int_\Sigma  \bd \sigma''\,2\, S(t,\sigma;t',\sigma'')\, \alpha^{t'}_{{}^{(2)}\check H}({}^{STT}e^{ab}(\sigma'')) \partial_a^{\sigma''}\partial_b^{\sigma''}S(t',\sigma'';0,\sigma')   
\nn\\
S'_{gr}(t,\sigma;0,\sigma')&:=&S(t,\sigma;0,\sigma')+ \nn \\
 && \int_0^t \bd t' \, \,   \int_\Sigma  \bd \sigma''\,2\, S(t,\sigma;t',\sigma'')\, \alpha^{t'}_{{}^{(2)}\check H}({}^{STT}e^{ab}(\sigma'')) \partial_a^{\sigma''}\partial_b^{\sigma''}S'(t',\sigma'';0,\sigma') \nn\\
\ea
such that
\ba\label{m6c}
{}^{(1)}\alpha_{\check H'}^{t}(\phi(\sigma))+{}^{(2)}\alpha_{\check H'}^{t}(\phi(\sigma))
&=& \int_\Sigma S_{gr}(t,\sigma;0,\sigma') \pi(\sigma')+S'_{gr}(t,\sigma;0,\sigma') \phi(\sigma') \, \bd \sigma' \q .\q\q
\ea
Finally we have to compute the second order complete observable associated to the first and second order propagated field. Note however that the second order (\ref{m6}) is already invariant under the constraints modulo terms of second order, since it is a sum of products of two phase space functions which are invariant to first order. Hence we only need to compute the second order term corresponding to the first order propagated field (\ref{m5}). According to (\ref{A22}) we have
\ba\label{m7}
 {}^{(2)}F_{[{}^{(1)}\alpha_{\check H'}^{t}(\phi(\sigma));\,T   ]} 
 &=&   \!\!\!
-\int_\Sigma  \bd \sigma' \big(  
S(t,\sigma;0,\sigma')  \partial_b (\pi{}^V T^b )(\sigma') +S'(t,\sigma;0,\sigma') (\partial_b \phi)(\sigma'){}^V T^b(\sigma') \big)  \nn\\
 && \!\!\!
 -  \int_\Sigma  \bd \sigma' \big(  \,
S(t,\sigma;0,\sigma') \partial_a ((\partial^a \phi){}^C T)(\sigma')  -m^2 \phi(\sigma') {}^C T(\sigma') +\nn\\
&& \q\q\q\;  S'(t,\sigma;0,\sigma') \pi(\sigma')
{}^CT(\sigma')\,\, \big)  \q\q .
\ea
The last term in (\ref{A22}) vanishes in this case, because in ${}^{(1)}\alpha_{\check H'}^{t}(\phi(\sigma))$ there only appear matter fields and no gravitational fields.

Now the second order complete observable is given by
\ba\label{m8}
{}^{[2]}F_{[\phi(\sigma);T]}(t)= {}^{(1)}\alpha^t_{\check H'}(\phi(\sigma))+{}^{(2)}\alpha^t_{\check H'}(\phi(\sigma))+ {}^{(2)}F_{[{}^{(1)}\alpha_{\check H'}^{t}(\phi(\sigma));\,T   ]} 
\ea
where the explicit expressions for the quantities on the right hand side can be found in equations (\ref{m5},\ref{m6},\ref{m7}). The first term in (\ref{m8}) coincides with the expression for a scalar field at time $t$ on a flat space--time. The other terms contain corrections due to the coupling to gravity: The second term ${}^{(2)}\alpha^t_{\check H'}(\phi(\sigma))$ is due to the fact, that the scalar field propagates on a space time with (non--interacting) gravitons, the last term ensures gauge invariance to second order.

To facilitate the interpretation of the result (\ref{m8}) note that the first two terms arise also if we evolve the scalar field $\phi$ with the time--dependent Hamiltonian
\ba\label{m9}
H(t)= \int_\Sigma \bd \sigma \frac{1}{2\gamma} ( \pi^2(\sigma) +(\delta^{ab}+ 2 \, {}^{STT}e^{ab}(t))\partial_a \phi \partial_b \phi (\sigma) +m^2 \phi^2 (\sigma))           
\ea
where the time dependence of ${}^{STT}e^{ab}(t)$ is given by 
\ba\label{m10}
{}^{STT}e^{ab}(t,\sigma) &=& \alpha^t_{{}^{(2)}\check H'}({}^{STT}e^{ab}(\sigma)) \nn\\
&=& \int_\Sigma  \left( {G'}^{ab}_{cd}(t,\sigma;0,\sigma'){}^{STT}a^{cd}(\sigma')+{G''}^{ab}_{cd}(t,\sigma;0,\sigma'){}^{STT}e^{cd}(\sigma')\right) \bd\sigma'  \; . \q\q\q
\ea

Here, if one evolves $\phi$ with the Hamiltonian (\ref{m9}) one does not treat the gravitational variables as dynamical anymore, i.e. the Poisson brackets between the gravitational variables are set to zero. Indeed (\ref{m9}) can be interpreted as the first order (in the graviton field) approximation of the Hamiltonian for a scalar field propagating on a graviton background. The first two terms in (\ref{m8}) are also the zeroth and first order approximation to the propagation of the scalar field with the time dependent Hamiltonian (\ref{m9}). Hence we captured in (\ref{m8}) the lowest order effect of a scalar field evolving on a graviton background. (The last term in (\ref{m8}) vanishes on the hypersurface, where all gravitational modes except for the $STT$--modes vanish.) Therefore the new propagator functions defined in (\ref{m6b}) are the zeroth and first order approximation to the propagator functions for a scalar field propagating on a graviton background. The higher order terms which arise if one evolves the scalar field with the Hamiltonian (\ref{m9}) can be found as  a subset of the higher order terms in the perturbative expansion of the complete observable associated to the scalar field.

\section{Poisson brackets: Space--time algebra of observables}\label{poisalg}

Here we will consider the Poisson brackets between two second order complete observables ${}^{[2]}F_{[\phi(\sigma_1);T]}(t_1)$ and ${}^{[2]}F_{[\phi(\sigma_2);T]}(t_2)$. Note that the Poisson bracket of two second order gauge invariants is an invariant of first order, we therefore need to consider only zeroth and first order terms of the Poisson bracket. The zeroth order will coincide with the result for the field commutator on a flat space--time, in particular it will vanish if $(t_1,\sigma_1)$ and $(t_2,\sigma_2)$ are space--like related (with respect to the Minkowski metric). The first order correction will be a function of the gravitational variables. Fluctuations in the gravitational variables will be reflected in fluctuations of the light cones, i.e. the causal structure.

The zeroth order of the Poisson bracket can be found to be
\ba\label{p1}
&& \{ {}^{(1)}\alpha^{t_1}_{\check H'}(\phi(\sigma_1)),\,{}^{(1)}\alpha^{t_2}_{\check H'}(\phi(\sigma_2))\} \nn\\
&& \q\q\q \q\q\q\q\q=
-\gamma                 
 \int_\Sigma \big( S(t_1,\sigma_1;0,\sigma')\,S'(t_2,\sigma_2;0,\sigma')-S'(t_1,\sigma_1;0,\sigma')\, S(t_2,\sigma_2;0,\sigma')
\big)\bd\sigma' \nn \\
&&\q\q\q\q\q\q\q\q     =-\gamma\, S(t_1,\sigma_1;t_2,\sigma_2)        \q .
\ea

For the first order of the Poisson bracket we have to consider the Poisson brackets between the first order term in (\ref{m8}) and the two second order terms in (\ref{m8}). To begin with we will show that one of the two contributions vanishes:
\ba\label{p2} 
 && \{{}^{(1)}\alpha^{t_1}_{\check H'}(\phi(\sigma_1)),\, {}^{(2)}F_{[{}^{(1)}\alpha_{\check H'}^{t_2}(\phi(\sigma_2));\,T   ]}  \} +  \{{}^{(2)}F_{[{}^{(1)}\alpha_{\check H'}^{t_1}(\phi(\sigma_1));\,T   ]} , \, {}^{(1)}\alpha^{t_2}_{\check H'}(\phi(\sigma_2)) \nn \\
&& \q\q\q \q\q\q\q\q\q\q\q=
\{ \,  {}^{(1)}\alpha^{t_1}_{\check H'}(\phi(\sigma_1)),\,  \int_\Sigma \{  {}^{(1)}\alpha_{\check H'}^{t_2}(\phi(\sigma_2)), {}^{[2]} \check C_K(\sigma') \} (-T^K(\sigma'))  \bd \sigma' \, \}
+  \nn\\
&& \q\q\q\q\q\q\q\q\q\q\q\q
\{ \, \int_\Sigma \{  {}^{(1)}\alpha_{\check H'}^{t_1}(\phi(\sigma_1)), {}^{[2]} \check C_K(\sigma') \} (-T^K(\sigma'))  \bd \sigma' \, ,\, 
 {}^{(1)}\alpha^{t_2}_{\check H'}(\phi(\sigma_2)) \}  
 \nn\\
&& \q\q\q\q\q\q\q \q\q\q\q=
\{ \,  {}^{(1)}\alpha^{t_1}_{\check H'}(\phi(\sigma_1)),\,  \int_\Sigma \{  {}^{(1)}\alpha_{\check H'}^{t_2}(\phi(\sigma_2)), {}^{[2]} \check C_K(\sigma') \} (-T^K(\sigma'))  \bd \sigma' \, \}
+  \nn\\
&& \q\q\q\q\q\q\q\q\q\q\q\q
\{ \, \int_\Sigma \{  {}^{(1)}\alpha_{\check H'}^{t_2}(\phi(\sigma_2)), {}^{[2]} \check C_K(\sigma') \} (-T^K(\sigma'))  \bd \sigma' \, ,\, 
 {}^{(1)}\alpha^{t_1}_{\check H'}(\phi(\sigma_1)) \}  \nn\\
&& \q\q\q\q\q\q\q\q\q\q\q = 0
\ea
where we used the Jacobi identity and the fact that the matter fields Poisson commute with the clock variables in the second equation.

The other contribution is
\ba\label{p3}
&& \{{}^{(1)}\alpha^{t_1}_{\check H'}(\phi(\sigma_1)),\,{}^{(2)}\alpha^{t_2}_{\check H'}(\phi(\sigma_2)) \} +
\{{}^{(2)}\alpha^{t_1}_{\check H'}(\phi(\sigma_1)),\,{}^{(1)}\alpha^{t_2}_{\check H'}(\phi(\sigma_2)) \}  \nn\\
&&= 
\{{}^{(1)}\alpha^{t_1}_{\check H'}(\phi(\sigma_1)),\, 
\int_0^{t_2} \bd t' \, \,  
 \int_\Sigma 2\, S(t_2,\sigma_2;t',\sigma')\,   \alpha_{{}^{(2)}\check H'}^{(t'} (     {}^{STT}e^{ab}(\sigma')) \,\partial_a^{\sigma'} \partial_b^{\sigma'}       \alpha_{{}^{(2)}\check H'}^{t'}     (\phi(\sigma')) \bd \sigma'   
    \,\,   \}  -  \nn\\
&& \q\q ((t_1,\sigma_1) \leftrightarrow (t_2,\sigma_2)) \nn\\
&&=
 -\gamma \int_0^{t_2} \bd t' \, \,  
 \int_\Sigma 2\, S(t_2,\sigma_2;t',\sigma')\,  
 \alpha_{{}^{(2)}\check H'}^{t'} (     {}^{STT}e^{ab}(\sigma')) \,\partial_a^{\sigma'} \partial_b^{\sigma'}       S(t_1,\sigma_1;t',\sigma')      \bd \sigma'  + \nn\\
&& \; \q    \gamma \int_0^{t_1} \bd t' \,\,\,   
 \int_\Sigma 2\, S(t_1,\sigma_1;t',\sigma')\,  
 \alpha_{{}^{(2)}\check H'}^{t'} (     {}^{STT}e^{ab}(\sigma')) \,\partial_a^{\sigma'} \partial_b^{\sigma'}       S(t_2,\sigma_2;t',\sigma')      \bd \sigma' 
\nn\\
&&=-\gamma \int_{t_2}^{t_1} \bd t' \,\,  
\int_\Sigma 2\, S(t_1,\sigma_1;t',\sigma')\,  
 \alpha_{{}^{(2)}\check H'}^{t'} (     {}^{STT}e^{ab}(\sigma')) \,\partial_a^{\sigma'} \partial_b^{\sigma'}       S(t',\sigma';t_2,\sigma_2)      \bd \sigma'  \q .
\ea
Comparing this result with the definition of the propagator function $S_{gr}$ in (\ref{m6b}) we see that for $t_2=0$ we can write
\ba\label{p4}
\{{}^{[2]}F_{[\phi(\sigma_1);T]}(t_1) \, , \, {}^{[2]}F_{[\phi(\sigma_2);T]}(t_1) \}=-\gamma S_{gr}(t_1,\sigma_1;0,\sigma_2) +O(2)  
\ea
where $O(2)$ refers to terms of second order. This result is similar to the Poisson bracket for a scalar field on flat space; the flat propagator function is replaced by the ``effective'' propagator function $S_{gr}$, which to the lowest non--trivial order takes into account the effects of the graviton background. Hence we can say that to this order the observables have the causality properties of field observables on such a (fixed) graviton background. Higher order terms of Poisson bracket between two complete observables will be much more complicated, for instance already the second order terms will include Poisson brackets between the (phase space dependent) propagator functions. Moreover the second order terms may include inverse spatial derivative operators making locality considerations quite difficult. 

However we can say, that all the higher order terms vanish in the zero gravity limit, that is for $\kappa,a_{ab},e_{ab} \rightarrow 0$. Indeed the higher order terms are all proportional to 
\ba\label{comp3}
\gamma \,\, (\text{grav. field})^m \,\, (\frac{\kappa}{\gamma} (\text{matter field})^2)^n
\ea    
for some powers $m,n \in \Nl$. We want to compare this behaviour to the case where one chooses four scalar fields as clocks for the scalar and vector constraints, see for instance \cite{bd2}. To this end we will consider the Poisson brackets restricted to the gauge fixing hypersurface, i.e. the hypersurface where the clock variables coincide with the parameters $\tau^K$. Furthermore we will consider only infinitesimally separated clock parameter values $\tau$ and $\tau+\epsilon$. Since we have \cite{bd1}
\ba
\{F_{[f;T]}(\tau),F_{[g;T]}(\tau)\}=F_{[\{f,g\}^*\,;\,T]}(\tau)
\ea 
where $\{\cdot,\cdot\}^*$ is the Dirac bracket with respect to the gauge $T^K(\sigma)\equiv\tau^K$ and \cite{bd1}
\ba
F_{[f;T]}(\tau+\epsilon)=  F_{[ f+\{f, \int \check C_K\epsilon^K \bd \sigma \}\,\,;T]}(\tau)+O(\epsilon^2)
\ea
we have to examine
\ba \label{comp1}
\{\phi(\sigma_1),\{\phi(\sigma_2), \int_\Sigma \epsilon^K \check C_K \bd \sigma \}\}^*  
&\simeq&
\{\phi(\sigma_1), \{\phi(\sigma_2), \int_\Sigma \epsilon^K \check C_K \bd \sigma \}\} \nn \\
&&+\int_\Sigma \{\phi(\sigma_1),T^L(\sigma') \}\{\check C_L(\sigma'),  \{\phi(\sigma_2), \int_\Sigma \epsilon^K \check C_K \bd \sigma \}\}  \nn\\
&&-\int_\Sigma \{\phi(\sigma_1), \check C_L(\sigma')\}\{T^L(\sigma'), \{\phi(\sigma_2), \int_\Sigma \epsilon^K \check C_K \bd \sigma \}\} \q ,
\ea
where we used the definition of the Dirac bracket, see for instance \cite{bd1}. We will be interested in the case where the field $\phi(\sigma)$ Poisson commutes with the clock variables, so that the second term in (\ref{comp1}) vanishes. Using the property of the Abelianized constraints, that $\{T^L(\sigma),\check C_K(\sigma')\}= \delta^L_K\delta(\sigma,\sigma')+O(C^2)$, one can see that also the third term vanishes:
\ba
\{T^L(\sigma'), \{\phi(\sigma_2), \int_\Sigma \epsilon^K \check C_K \bd \sigma \}\} = \{\phi(\sigma_2), \{T^L(\sigma') , \int_\Sigma \epsilon^K \check C_K \bd \sigma \}\} \simeq 0 \q .
\ea
With the definition (\ref{d2}) of the constraints $\check C_K$
\ba
\check C_K(\sigma) &=&\int_\Sigma C_j(\sigma') (A^{-1})^j_K(\sigma,\sigma) \bd\sigma' +\int_\Sigma \mu_K^{L_1L_2}(\sigma,\sigma',\sigma'')\tilde C_{L_1}(\sigma')\tilde C_{L_2}(\sigma'')\bd\sigma'\bd\sigma'' +O(C^3) \nn\\
\mu_K^{L_1L_2}(\sigma,\sigma',\sigma'')&=& -\frac{1}{2} \{T^{L_1}(\sigma'),\{T^{L_2}(\sigma''),\tilde C_K(\sigma)\}\}
\ea
the first term is equal to
\ba\label{z1}
\{\phi(\sigma_1), \{\phi(\sigma_2), \int_\Sigma \epsilon^K \check C_K \bd \sigma \}\} &\simeq&
\int_\Sigma \epsilon^K(\sigma)\, \{ \phi(\sigma_1),\{\phi(\sigma_2),C_j(\sigma')\}\} \, (A^{-1})^j_K(\sigma',\sigma)\, \bd\sigma\bd\sigma' +
\nn \\ &&
 \int_\Sigma \epsilon^K(\sigma) \{\phi(\sigma_1), (A^{-1})^j_K(\sigma',\sigma)\}\,\,\{\phi(\sigma_2), C_j(\sigma')\}\, \bd\sigma\bd\sigma' +
\nn \\ &&
\int_\Sigma \epsilon^K(\sigma) \{\phi(\sigma_2), (A^{-1})^j_K(\sigma',\sigma)\}\,\,\{\phi(\sigma_1), C_j(\sigma')\}\, \bd\sigma\bd\sigma' +
\nn\\
&&
\int_\Sigma   \epsilon^K(\sigma)   \{\phi(\sigma_1),\tilde C_{L_1}(\sigma') \} \{\phi(\sigma_2),\tilde C_{L_2}(\sigma'')\} \times \nn\\ 
&& \q
 \{T^{L_1}(\sigma'),\{T^{L_2}(\sigma''),C_l(\sigma''')\}\}(A^{-1})^{l}_K(\sigma''',\sigma)   \,  \bd\sigma\bd\sigma'\bd\sigma''\bd\sigma'''   \q\,  . \q\q
\ea

We will apply this formula to the case were one uses four scalar fields $S^K(\sigma)$ as clock variables. We have to add the following terms to the scalar and vector constraints:
\ba
{}^{S}C &=& \frac{1}{2 \alpha}\sum_{K=0}^3 ( (\Pi_{K})^2 + q q^{ab}\partial_a S^K \partial_b S^K +q V(S^K)) \nn \\
{}^SV_b &=&\frac{1}{\alpha} \sum_{K=0}^3  \, \Pi_K \partial_b S^K
\ea
where $\Pi_K$ is the conjugated momentum to $S^K$ with $\{S^K,\Pi_L\}=\alpha\delta^K_L$ and $V(S^K)$ is a potential for the scalar field $S^K$. Then the second and third term in (\ref{z1}) vanishe and we are left with
\ba\label{z2}
\{\phi(\sigma_1), \{\phi(\sigma_2), \int_\Sigma \epsilon^K \check C_K \bd \sigma \}\}
 &\simeq&
\gamma \delta(\sigma_1,\sigma_2) \int_\Sigma \epsilon^K(\sigma) (A^{-1})^0_K(\sigma_1,\sigma)\bd\sigma +
\nn \\ &&
\alpha\int_\Sigma   \epsilon^K(\sigma)   \{\phi(\sigma_1),\tilde C_{L_1}(\sigma') \} \{\phi(\sigma_2),\tilde C_{L_2}(\sigma')\} \delta^{L_1L_2} (A^{-1})^{0}_K(\sigma',\sigma) \bd\sigma\bd\sigma' \nn\\
\ea
Here $\epsilon'(\sigma'):= \int \epsilon^K(\sigma)(A^{-1})^{0}_K(\sigma',\sigma) \bd \sigma $ is the translation to the factor that would arise if we would use $\int \epsilon'(\sigma') C(\sigma')\bd\sigma'$ as time generator. We therefore see 
\ba\label{z3}
\gamma \frac{\alpha}{\gamma} \int \epsilon'(\sigma')\{\phi(\sigma_1),\tilde C_{L_1}(\sigma') \} \{\phi(\sigma_2),\tilde C_{L_2}(\sigma')\} \delta^{L_1L_2} \bd\sigma' &\simeq & \gamma \frac{\alpha}{\gamma} \int \epsilon'(\sigma') \{\phi(\sigma_1),C_{j}(\sigma')\}\{\phi(\sigma_2),C_{k}(\sigma')\}  \times \nn \\
&&\q\q
 (A^{-1})^j_{L_1}(\sigma')(A^{-1})^k_{L_2}(\sigma')\delta^{L_1L_2}\bd\sigma'
\ea
as the genuine effect of this choice of clock variables on the Poisson brackets. In (\ref{z3}) we used that the inverse matrix $A^{-1}$ can be written as $ (A^{-1})^j_{L_1}(\sigma',\sigma'')= (A^{-1})^j_{L_1}(\sigma')\delta(\sigma',\sigma'')$ if one uses scalar fields as clock variables. Since $A^K_j(\sigma,\sigma')=\{T^K(\sigma),C_j(\sigma')\}$ the term (\ref{z3}) can be interpreted as the generalization of the expression ``energy of the observed field divided by the energy of the clock variables'' which we derived for the case of parametrized particles in the introduction, section \ref{intro}. The correction (\ref{z3}) can be made small by choosing $\alpha$ to be very small (or equivalently the energy of the clock variables very large). But one has to keep in mind that through the coupling to gravity backreaction terms arise which scale with (positive powers of) $\kappa/\alpha$, hence one has to balance between the term in (\ref{z3}) and the backreaction terms. In contrast to this result, the corrections to the Poisson bracket in the case of the ADM clocks scale in the same way as the backreaction terms. Another point is that using the ADM clocks one can perturb around flat space, whereas if one uses scalar fields one has rather to perturb around a phase space point with a finite energy density and hence a non--vanishing gravitational field due to the constraint equations.

\section{Summary and discussion}

In this work we introduced a perturbation scheme for the calculation of Dirac observables. For this perturbation scheme one has to choose a fixed phase space point providing a background space--time. Dirac observables can be calculated order by order in the fluctuations around this phase space point.

We applied this method to general relativity and chose as the fixed phase space point the Minkowski background. However the method is also applicable to other backgrounds, for instance a cosmological background. Furthermore one can also choose another set of clock variables, as long as these clocks provide a good parametrization of the gauge orbits near the fixed phase space point. Our choice of the ADM clocks was guided by the aim to obtain observables which approximate very well the field observables on a flat background.

The first order approximation of the complete observables coincides with the observables of the linearized theory, which gives a precise understanding of how the observables and the dynamics of the linearized theory, as for instant the graviton are connected to the observables of the full theory. 

The second order terms of the complete observables associated to matter fields take into account the propagation of these matter fields on a graviton background, that is the scattering of matter from gravitons, the third order term will contain among other things backreaction terms.\footnote{If we would rescale our gravitational variables to $a'_{ab}:=\kappa^{-1/2}a_{ab}$ and  $e'_{ab}:=\kappa^{-1/2}e_{ab}$ the first order of a complete observable associated to a matter field is proportional to $\kappa^0$ and in general the $m$-th order to $\kappa^{(m-1)/2}$.}

Formula (\ref{A14}) and (\ref{A17}) provide two different ways to calculate these Dirac observables. In (\ref{A14}) we first time evolve the (gauge variant) field in question and afterwards calculate the gauge invariant extension of the result. It is shown in section \ref{propa} that as the generator for the time evolution we can choose the function $\check H'$, which does only contain the ``physical modes'' of the linearized theory, i.e. only matter fields and the symmetric transverse traceless $(STT)$ modes of the gravitational field. This facilitates the interpretation of the resulting time evolution as one involving the dynamics and the scattering of gravitons (the $STT$--modes). We have different interaction processes contributing to the time evolution and for each individual interaction process one could in principle calculate the fully gauge invariant extension. In this way one can associate a Dirac observable to each interaction process.

On the other hand in (\ref{A17}) we first calculate the complete observable or gauge invariant extension associated to the field in question and then evolve the resulting observable with a generating function which is also a Dirac observable and is given by the ADM energy. This shows that our time evolution corresponds to time translations at infinity and that with our choice of time we have a Hamiltonian whose matter part coincides with the matter part of the Hamiltonian constraint. Note however that this time generating function is not unique -- one can add an arbitrary combination of the constraints without changing the result on the constraint hypersurface. For instance one can choose between the ADM energy, which in our case is of finite order and the function $\check H$ which includes infinitely high order terms. 

The resulting Dirac observables have a local interpretation stemming from the flat space--time limit, they give the fields at some space time point coordinatized by $(t,\sigma)$. The interpretation of the coordinates $(t,\sigma)$ is not as straightforward as in the case where one uses matter fields as clock variables: what one can say is that the measurement of these observables has to be performed with respect to a reference system which is as near to the Minkowskian one as possible. Thus to zeroth order the proper distance between two points can be obtained by using the flat metric, for higher orders one has to use the (time evolved) gravitational field $e_{ab}(t,\sigma)$. However a better understanding of the geometrical meaning of these coordinates would be helpful. An interesting question is whether the  ADM clocks can be understood as an approximation to a set of clocks which has a more obvious geometrical interpretation. 


Furthermore we suggested to consider the Poisson algebra of the complete observables corresponding to different space--time points, in order to learn more about the locality properties and the interpretation of the complete observables. We made some preliminary steps into this direction and calculated the first order of the Poisson bracket. The first order approximation of the Poisson bracket (\ref{p4}) can be interpreted to reflect the causality properties of a space--time with (non--interacting) gravitons. Hence to this order one can say that the complete observables with this choice of clocks represent a local measurement. But higher orders will include inverse derivative operators stemming from our choice of clock variables. In particular our choice of time parameter $t$ is very global since it corresponds to time translations at asymptotic infinity. It may help to define another one--parameter family of clock value parameters $\tau^K(t')$, for which the $\tau^K$ change only in a finite region, effectively introducing a boundary in space--time.

Choosing scalar fields as clock variables will lead to a better locality behaviour, since the complete observables associated to these clocks will not involve inverse derivatives, see also \cite{bd2} for a discussion of the advantages using scalar fields. However we showed that the Poisson brackets between (infinitesimally time--separated) complete observables and hence the uncertainty relations between these complete observables include a term which is inverse to the energy of the clock fields. If one wants to keep this term small one has to increase the energy of the clock fields which leads to larger backreactions and hence to a bound from below for the additional term in the Poisson bracket.

In summary we think that the approximation scheme introduced here can be very useful to calculate Dirac observables and to explore their properties. A better understanding of the properties of complete observables is needed, in particular of the question how the choice of clock variables influences these properties.

\begin{appendix}

\section{Tensor modes} \label{tensor modes}

Similar to the longitudinal and transversal modes for a vector field on $\Rl^3$ one can introduce tensor modes for a tensor field. For a proof of the completeness of these modes, see \cite{mcp4}. To begin with we define the projector onto the transversal modes of a vector field by
\be
(p\cdot v)_a:=p_a^b\cdot v_b:=(\delta_a^b+W^{-2}\cdot \partial_a \partial_b)\cdot v_b \q .
\ee
This allows us to introduce the following basis of tensor modes:
\begin{xalignat}{2}\label{lingravBasis2}
&({}^{LT}P\cdot T)_{ab}=(\delta_a^c-p_a^c)\cdot p_b^d \cdot T_{cd} && \mbox{2 left long. right transv. modes}\nn \\
&({}^{LL}P\cdot T)_{ab}=(\delta_a^c-p_a^c)\cdot(\delta_b^d-p_b^d) \cdot T_{cd} && \mbox{1 left and right long. mode}\nn \\
&({}^{TL}P\cdot T)_{ab}=p_a^c\cdot (\delta_b^d-p_b^d) \cdot T_{cd} &&\mbox{2 left transv. right long. modes}\nn \\
&({}^{T}P\cdot T)_{ab}= \tfrac{1}{2}p_{ab}\cdot p^{cd}\cdot T_{cd} && \mbox{1 symm. transv. trace part mode} \nn \\
&({}^{AT}P\cdot T)_{ab}=\tfrac{1}{2}(p_a^c\cdot p_b^d-p_b^c\cdot p_a^d)\cdot T_{cd} &&\mbox{1 antisymm. transv. mode} \nn \\
&({}^{STT}P\cdot T)_{ab}= \tfrac{1}{2}(p_a^c\cdot p_b^d +p_b^c\cdot p_a^d-p_{ab}p^{cd})\cdot T_{cd} &&\mbox{2 symm. transv. tracefree modes}
\end{xalignat}
 Using the projector property $p\cdot p=p$, it is easy to see that the 
 projectors ${}^{X}P$ are orthogonal to each other and satisfy
 ${}^{X}P\cdot{}^{Y} P=\delta^{XY}\,\,{}^{X}P$. Furthermore the set of projectors is complete, that is
\ba
\sum_X {}^{X}P_{ab}^{cd}=\delta^c_a \delta^d_b  \q .
\ea

\section{First order of the matrix A} \label{FA}

Here we display the relevant terms to find the third order of the time generator $\check H'$ used in section \ref{propa}:
\ba\label{app1}
\{{}^GT^a(\sigma),\,{}^{(2)}C[1]\}
&=&
W^{-2}\beta(-2\epsilon^{aef}\partial_e\partial^g+3 \epsilon^{gef}\partial_e\partial^a)a_{fg}
\nn\\
\{{}^VT^a(\sigma),\,{}^{(2)}C[1]\}
&=&
2\beta (\, (W^{-4}\partial^a\partial_c\partial_e +W^{-2}\partial_c\delta^a_e+W^{-2}\partial_e \delta^a_c ) a^{ce} +  
\nn \\
&& \q\;\; \beta W^{-2}(\partial^d\partial_b \epsilon^{abc}-\partial^a\partial_b \epsilon^{dbc}   )e_{dc} \,
)
\nn\\
\{{}^CT(\sigma),\,{}^{(2)}C[1]\}
&=&
-\frac{1}{2}W^{-2}\partial_a \partial^e {e^a}_e+ \frac{1}{2}W^{-2} \partial_c\partial^c {e^a}_a -\beta W^{-2} \epsilon^{efd}\partial_e a_{fd} 
\ea
The right hand sides of the equations (\ref{app1}) do not contain $STT$--modes, which shows that the third order of $\check H'$ cn be obtained by just restricting the third order of $C[1]$ to the $STT$--modes.

\section{Propagators}\label{propagator}

Here we will review the propagators for the scalar field and the graviton on a flat background. We will start with the scalar field. Given initial values $\phi(\sigma)$ and $\pi(\sigma)$  on $\Sigma$ the scalar field $\phi(t,\sigma)$ at a later time $t$ can be calculated to be
\ba\label{aa1}
\phi(t,\sigma) &=&\sum_{r=0}^\infty\, \{\phi(\sigma), {}^{(2)}{}^{\phi}C[1]\}_r\, \frac{t^r}{r!}  \nn\\
               &=& \cos\big[ (-\Delta_\sigma +m^2)^{1/2} t \big] \phi(\sigma)+ (-\Delta_\sigma+m^2)^{-1/2}\sin\big[(-\Delta_\sigma +m^2)^{1/2} t] \pi(\sigma)
\ea
where ${}^{(2)}{}^{\phi}C[1]=\int_\Sigma \frac{1}{2}(\pi^2+\partial_a \phi \partial^a \phi +m^2 \phi^2) $ and $\Delta=\delta^{ab}\partial_a\partial_b$. Introducing a delta function this can be rewritten as
\ba\label{aa2}
\phi(t,\sigma)&=& \int_\Sigma  \cos\big[ (-\Delta_\sigma +m^2)^{1/2} t \big]\delta(\sigma,\sigma') \phi(\sigma')+ (-\Delta_\sigma+m^2)^{-1/2}\sin\big[(-\Delta_\sigma +m^2)^{1/2} t] \delta(\sigma,\sigma')\pi(\sigma') \bd \sigma' \nn\\
&=:& \int_\Sigma   S'(t,\sigma;0,\sigma') \phi(\sigma')+ S(t,\sigma;0,\sigma') \pi(\sigma)\, \bd \sigma' \q .
\ea
From the last equation one can read off the propagator functions $S$ and $S'$. This definition of the propagator function can be generalized by
\ba\label{aa3}
S(t_1,\sigma_1;t_2,\sigma_2):=S(t_1-t_2,\sigma_1;0,\sigma_2)
\ea
and the analogous definition for $S'$. We have that $S(t_1,\sigma_1;t_2,\sigma_2)=-S(t_2,\sigma_2;t_1,\sigma_1)$ is odd under permutation of $t_1,t_2$, whereas $S'(t_1,\sigma_1;t_2,\sigma_2)=S'(t_2,\sigma_2;t_1,\sigma_1)$ is even.

In the same way one can determine the propagator functions for the graviton field. The Poisson brackets of the graviton fields ${}^{STT}e_{ab}={}^{STT}P^{cd}_{ab}e_{cd}$ and ${}^{STT}a_{ab}$ with the second order Hamiltonian 
\ba
{}^{(2)}\check H'= \frac{1}{\kappa}\int_\Sigma ( 2\beta \epsilon^{bed}\,\, {}^{STT}{e^a}_e\,\partial_b \,\,{}^{STT}a_{ad}-\beta^2 \,\,{}^{STT}{a}^{ab}\,\,{}^{STT}{a_{ab}} )\,\bd \sigma  
\ea
 defined in section \ref{propa} are given by 
\ba \label{aa4}
\{{}^{STT}a_{ab}(\sigma),{}^{(2)}\check H'\} 
&=& 2 \beta D^{fe}_{ab}\,\,{}^{STT}a_{fe}(\sigma)
:=2 \beta \frac{1}{2}\partial_c\epsilon^{cde}(\delta_{db}\delta^f_a+
\delta_{da}\delta^f_b)\,\,{}^{STT}a_{fe}(\sigma) \nn\\
\{{}^{STT}e_{ab}(\sigma),{}^{(2)}\check H'\} 
&=& 
2\beta^2 \,\,{}^{STT}a_{ab}(\sigma)-2 \beta D^{fe}_{ab}\,\,{}^{STT}e_{fe}(\sigma)  \nn\\
\{{}^{STT}e_{ab}(\sigma),{}^{(2)}\check H'\}_2 
&=&
[(2\beta D)^2]^{fe}_{ab}\,\,\, {}^{STT}e_{fe}(\sigma)
\q .
 \ea
Hence the graviton fields are evolved according to
\ba \label{aa5}
{}^{STT}a(t,\sigma)&=& \int_\Sigma {G}_{ab}^{fe}(t,\sigma; 0, \sigma')\,\, {}^{STT}a_{fe}(\sigma') \bd \sigma'  \nn\\
{}^{STT}e(t,\sigma)&=&\int_\Sigma {G'}_{ab}^{fe}(t,\sigma; 0, \sigma')\,\, {}^{STT}a_{fe}(\sigma') + {G''}_{ab}^{fe}(t,\sigma; 0, \sigma')\,\, {}^{STT}e_{fe}(\sigma')      \bd \sigma' 
\ea
with
\ba
{G}_{ab}^{fe}(t,\sigma; 0, \sigma') &=& \big[\exp(2\beta t D_\sigma)\big]_{ab}^{fe} \delta(\sigma,\sigma') \nn\\
{G'}_{ab}^{fe}(t,\sigma; 0, \sigma') &=& 2\beta^2\sum_{r=0}^\infty \frac{t^{2r+1}}{(2r+1)!} [(2\beta D_\sigma)^{2r}]^{fe}_{ab} \delta(\sigma,\sigma') \nn\\
{G''}_{ab}^{fe}(t,\sigma; 0, \sigma') &=& \big[\exp(-2\beta t D_\sigma)\big]_{ab}^{fe} \delta(\sigma,\sigma')   \q .
\ea

\end{appendix}

\vspace{2cm}
~\\
{\large{\bf Acknowledgements}}
\vspace{0.2cm}
~\\
We are grateful to Thomas Thiemann for discussions, BD thanks furthermore Benjamin Bahr, Lee Smolin and Simone Speziale for discussions. We used the computer algebra system Cadabra \cite{cadabra} and JT thanks Kasper Peeters for correspondence. 
JT thanks furthermore the German National Merit Foundation for financial support and the Perimeter Institute for Theoretical Physics for hospitality. This work was supported in part by the Government of Canada through NSERC and by the Province of Ontario through MEDT.

\end{document}